\documentclass[aps,pre,twocolumn,groupedaddress]{revtex4}
\usepackage{amssymb}
\usepackage{amsmath}
\usepackage{cases}
\usepackage{graphicx,color}
\definecolor{r}{rgb}{1,0,0}   
\definecolor{g}{rgb}{0,1,0}   
\definecolor{b}{rgb}{0,0,1}

\begin{document}


\title{Hyperuniformity Disorder Length Spectroscopy for Extended Particles}


\author{D. J. Durian}
\affiliation{
     Department of Physics \& Astronomy, University of
     Pennsylvania, Philadelphia, PA 19104, USA
}


\date{\today}

\begin{abstract}
The concept of a hyperuniformity disorder length $h$ was recently introduced for analyzing volume fraction fluctuations for a set of measuring windows [Chieco {\it et al.} (2017)].  This length permits a direct connection to the nature of disorder in the spatial configuration of the particles, and provides a way to diagnose the degree of hyperuniformity in terms of the scaling of $h$ and its value in comparison with established bounds.  Here, this approach is generalized for extended particles, which are larger than the image resolution and can lie partially inside and partially outside the measuring windows.  The starting point is an expression for the relative volume fraction variance in terms of four distinct volumes: that of the particle, the measuring window, the mean-squared overlap between particle and region, and the region over which particles have non-zero overlap with the measuring window.  After establishing limiting behaviors for the relative variance, computational methods are developed for both continuum and pixelated particles.  Exact results are presented for particles of special shape, and for measuring windows of special shape, for which the equations are tractable.  Comparison is made for other particle shapes, using simulated Poisson patterns.  And the effects of polydispersity and image errors are discussed.  For small measuring windows, both particle shape and spatial arrangement affect the form of the variance.  For large regions, the variance scaling depends only on arrangement but particle shape sets the numerical proportionality.  The combined understanding permit the measured variance to be translated to the spectrum of hyperuniformity lengths versus region size, as the quantifier of spatial arrangement.  This program is demonstrated for a system of non-overlapping particles at a series of increasing packing fractions as well as for an Einstein pattern of particles with several different extended shapes.
\end{abstract}

\pacs{05.40.-a, 46.65.+g, 82.70.-y}


\maketitle





The structural uniformity of a many-body system may be studied in terms of fluctuations in the number \cite{TorquatoPRE2003, GabrielliPRD2002} and volume fraction \cite{ZacharyJSM2009, BerthierPRL2011, ZacharyPRL2011} of objects inside measuring windows of equal size but different locations.  At one extreme, a totally random arrangement exhibits large fluctuations that are Poissonian, such that the volume fraction variance for large $L$ scales as ${\sigma_\phi}^2(L)\sim 1/L^d$, where $L$ is the width of the measuring windows and $d$ is dimensionality.  By contrast, a  ``hyperuniform" \cite{TorquatoPRE2003}  or ``superhomogeneous''  \cite{GabrielliPRD2002} arrangement exhibits smaller sub-Poissonian fluctuations that decay more rapidly as ${\sigma_\phi}^2(L)\sim 1/L^{d+\epsilon}$ with $0<\epsilon \le1$.  At the $\epsilon=1$ extreme, two straightforward hyperuniform arrangements are ``shuffled lattice" \cite{GabrielliPRD2002} and ``Einstein'' \cite{ATCpixel} patterns, where particles are effectively bound by square-well and harmonic potentials, respectively, to fixed crystalline lattice sites and are independently displaced as though by thermal energy.   If the root mean square displacement is large compared to the lattice spacing, then the arrangement appears quite random to the eye.  In such cases, the underlying crystalline order is well hidden. 

There has been growing interest in hyperuniformity because it occurs jointly with the existence of special materials properties.  Examples include jamming in amorphous materials \cite{DonevPRL2005, BerthierPRL2011, ZacharyPRL2011, KuritaPRE2011, DreyfusPRE2015, WuPRE2015}, complete optical band gaps in disordered photonics materials \cite{FlorescuPNAS2009, ManPNAS2013, MullerAOM2014, ManOE16, Scheffold2016}, and reversibility/irreversibility in periodically driven systems \cite{HexnerPRL2015, TjhunPRL2015, WeijsPRL2015}.  Hyperunformity is also important in the arrangement of photoreceptors in the retina \cite{JiaoPRE2014}, and in the large-scale structure of the universe \cite{GabrielliPRD2002}.  Unfortunately, hyperuniformity can be quite delicate to diagnose \cite{DreyfusPRE2015}.  Clean power-law behavior, over many decades, are needed to convincingly establish the value of $\epsilon$ and determine if it is nonzero.  For polydisperse systems, the signature of hyperuniformity is absent in number but not volume-fraction fluctuations.  Plus, as shown here, particle {\it shape} imparts systematic features to the functional form of ${\sigma_\phi}^2(L)$ that have nothing to do with spatial arrangement of the particles, and that can extend out to many times the particle width.  These difficulties are compounded by finite size effects: Measurements of ${\sigma_\phi}^2(L)$ must incorrectly decreasing to zero as $L$ approaches system size because allowed measuring windows all strongly overlap and contain the same particles.  In reciprocal space, finite size errors in the spectral density are not so dramatic, being statistical rather than systematic \cite{AtkinsonPRE2016, ST}.  

To help diagnose the uniformity of particle arrangements we recently introduced a ``hyperuniformity disorder length", $h(L)$, that can be extracted from ${\sigma_\phi}^2(L)$ data \cite{ATCjam, ATCpixel}.  This was done in the context of both point and  ``pixel" particles, whose width equals the resolution/precision $p_o$ of the experiment/simulation.  This is convenient for actual data coming from digital cameras, where $p_o$ is the pixel width; however, preprocessing is required to identify each particle and set the value of its central pixel to particle volume divided by voxel volume, ${p_o}^d$.  The intuitive meaning of $h$ is to specify the distance from the boundary of the measuring windows over which fluctuations are important.  It scales as $h\sim L$ for Poissonian arrangements, where a fixed fraction of the entire volume is important, and is constant for strongly hyperuniform arrangements with $\epsilon=1$.  For Einstein patterns, the asymptotic value is about half the root mean square displacement.  In general, for ${\sigma_\phi}^2\sim L^{d+\epsilon}$ the scaling is $h\sim L^{1-\epsilon}$ and value of $h$ is bound by $L/2$ as the upper limit for a totally random arrangement.  Thus the {\it value} of $h(L)$ as well as the form of $h(L)$ versus $L$ have direct meaning, both physically and in comparison with the bounds.  By contrast, prior uses of ${\sigma_\phi}^2(L)$ and the spectral density $\chi(q)$, the reciprocal space analogue, for disordered systems focus on scaling behavior and make no use of the actual values of ${\sigma_\phi}^2(L)$ and $\chi(q)$.

In this paper we develop the formalism for finding the real-space spectrum $h(L)$ for experimental or simulated arrangements of extended particles, which have nonzero volume $V_P$ larger than the voxel volume ${p_o}^d$, and hence spread across multiple pixels.  One goal is to enable application of  ``hyperuniformity disorder length spectroscopy" (HUDLS) to digital video data, directly, without the preprocessing steps of finding the positions and volumes of every particle and creating a corresponding pixelated image.  The key ingredient is prediction of the volume fraction variance for a totally random arrangement of the same objects.  This was relatively straightforward for pixel particles \cite{ATCpixel}.  By contrast, as our main topic, it is more complicated for extended particles because they may lie partially inside and partially outside a measuring window.

We begin by discussing how to extract ${\sigma_\phi}^2(L)$ and its statistical uncertainty from image data.  Then we show how to compute the variance function for totally random particle arrangements, first for pixel particles as review and then for extended particles as a new result.  This is done for both pixelated and continuum particles of arbitrary shape and size.  After developing the general methods and examining special limits, we evaluate the variance function for particles of various specific shapes and we demonstrate the validity of the predictions by analysis of simulated 2-dimensional random arrangements.  Lastly, as a small demonstration, we analyze $h(L)$ for two types of non-random arrangements of extended particles: non-overlapping particles, and Einstein patterns.


\section{Image processing}

The primary measurable is the variance ${\sigma_\phi}^2(L)$ for fluctuations in the volume fraction $\phi$ occupied by particles inside measuring windows of volume $V_\Omega\propto L^d$ placed throughout a $d$-dimensional image.  Here $L$ represents the width of the window, e.g.\ the side length of a hypercubic window or the diameter of a hyperspherical window.  Other window shapes are possible, and can be implemented using a dimensionless window function $H(x,y,z,\ldots;L)$ whose integral over space equals the window volume $V_\Omega$.  Usually $H$ is taken as a step function (1 inside and 0 outside), but a Gaussian or Lorentzian etc.\ could also be used to help smooth out noise in experimental data.  Raw image data from simulation or experiment consist of binary or grayscale ``intensity'' values $I(x,y,z\ldots)$ where the coordinates specify the location of the cubic pixels, or voxels, of side-length $p_o$ and volume ${p_o}^d$.  Ideally, images are normalized such that the volume of a particle is $V_P=\sum I {p_o}^d$, where the sum is over the pixels covered by the particle.  Then the volume fraction for a particular measuring window equals the sum of intensity values divided by the number of pixels in the window.  This is to be computed for many window locations, from which the average and variance are to be found.  The average is just the volume fraction $\phi$ of the entire sample.  The variance ${\sigma_\phi}^2(L)$ depends on window size and the nature of the particle arrangement; it is the key quantity to be analyzed per the following sections.  For this we define a relative variance as
\begin{equation}
	{\mathcal V}_{data}(L) = \frac{ {\sigma_\phi}^2(L) }{\phi}.
\label{relvar}
\end{equation}
With this normalization by $\phi$, the relative variance will be seen to have a large-window asymptote of $\mathcal V\rightarrow \langle H^2\rangle V_P/V_\Omega$ for random (Poisson) arrangements of particles of volume $V_P$, no matter what the volume fraction or particle shape or window shape.  In general, the window shape does not affect scaling

The standard procedure is to compute ${\sigma_\phi}^2(L)$ from the list of volume fractions for a large number of random locations for a measuring window of a given size \cite{TorquatoPRE2003}.  Here, instead, we use a Fourier technique to compute the list of volume fractions for {\it all} possible locations of the measuring window.  The basis for this is that the sum of intensity values in a given window equals the convolution of the image with the window function $H$.  Therefore, by the convolution theorem, $\mathcal{F}^{-1}[ \mathcal{F}(H)\mathcal{F}(I)]$ is a $d$-dimensional matrix where each entry is the sum of intensity values for a window at a location specified by the indices.  For pixelated image data, $\mathcal{F}$ is the discrete Fourier transform as implemented for example in Mathematica by the {\tt Fourier} function.  While we present the formalism in general and give calculations for several specific particle- and window shapes, all simulation tests are for two-dimensional systems and square $L\times L$ step-function windows with $H$ constructed of 0s and 1s.

For samples of finite size it's important to understand the possible errors that can arise.  The statistical uncertainty of the volume fraction variance may be estimated as
\begin{equation}
	\Delta{\sigma_\phi}^2={\sigma_\phi}^2\sqrt{2/(s-1)},
\label{DeltaSigma}
\end{equation}
where $s$ is the number of independent samples.  For the Fourier method, $s$ is simply the ratio of image volume to window volume.  To our knowledge statistical uncertainty was not estimated prior to Ref.~\cite{ATCpixel}, where the volume fraction variance was found by standard procedure and $s$ was estimated as the volume of the image that was covered by the randomly-chosen set of measuring windows, divided by window volume.  Ref.~\cite{ATCpixel} also shows how to estimate the statistical uncertainty for small windows, where the measured distributions are not Gaussian.  It is important to note that Eq.~(\ref{DeltaSigma}) does not represent the degree of noise, i.e.\ the smoothness, of ${\sigma_\phi}^2(L)$ versus $L$ results for one image.  Indeed a given spectrum is perfectly smooth for the Fourier method since all possible window locations are samples.  Noise in ${\sigma_\phi}^2(L)$ versus $L$ only arises when the image is undersampled in the traditional method of placing, say, $10^4$ windows at random locations.  What Eq.~(\ref{DeltaSigma}) truly represents is the scatter of ${\sigma_\phi}^2(L)$ at a given $L$ for an ensemble of statistically equivalent patterns.

This Fourier method has several advantages over the usual procedure.  First, it is simple to implement, especially for cubic windows.  It is fast.  It gives better statistics, since {\it all} possible window locations are used.  It allows for easy estimation of $s$ in computing the statistical uncertainty of the variance.  And it can be implemented for extremely large systems, if small-scale features are not of interest, by suitable coarse-graining.

Prior to variance computation, it is important to correct for experimental data that are unnormalized or that have known artifacts.   If the light illumination/collection fields are flat, then the measured digital image data for bubbles, colloids, grains, etc.\  may be written $I_{m}=\alpha I + I_{a}+I_{g}$ where $I$ is the true signal, $\alpha$ is a normalization factor, $I_{a}$ is an additive constant, and $I_{g}$ is a Gaussian random variable of zero mean.  Then the true volume fraction and variance are $\phi=(\phi_m-\phi_{a})/\alpha$ and ${\sigma_\phi}^2(L) = ( {\sigma_m}^2 - n_B{\sigma_{g}}^2 )/\alpha^2$, where $n_B=(L/p_o)^d$ is the number of pixels in the window.


\section{Pixel Particles}

We begin by recalling the results of Ref.~\cite{ATCpixel} for arrangements of a mixture of different pixel particle species with ``volumes" $V_{P_i}=I_i {p_o}^d$.  Such a pixel pattern could represent point particles, or it could be a ``central pixel'' representation of extended particles of actual volume $V_{P_i}$.  In either case, the intensity of each pixel is incremented by $+I_i$ for each particle of species $i$ that is centered upon it, and the volume fraction equals the average intensity per pixel.  For random ``multinomial'' pixel patterns, the intensity of each pixel is randomly drawn from $\{0, I_1, I_2,\ldots\}$ with some set of probabilities $\{1-\sum q_i, q_1, q_2,\ldots\}$.  Then only one particle at a time resides on each pixel, and the relative variance for cubic measuring windows of volume $V_\Omega=L^d$ was computed in Ref.~\cite{ATCpixel} to be
\begin{equation}
	 \frac{ {\sigma_\phi}^2(L) }{\phi} = \left( 1 - \frac{\phi}{\langle I_P \rangle}\right) \frac{ \langle I_P \rangle {p_o}^d }{L^d},
\label{vpixmulti}
\end{equation}
where $\langle I_P\rangle = \sum \phi_i I_i / \phi$ is the volume-fraction weighted average particle intensity.  For random ``Poisson" pixel patterns, particles are placed at random -- including on top of each other, and the relative variance was instead found to be ${\sigma_\phi}^2(L)/\phi = \langle I_P \rangle {p_o}^d/L^d$.  This same result holds for a random multinomial pattern if particle volumes are all large, such that the intensity $I_i=V_{P_i}/{p_o}^d$ is large and the average probability $q_i=\rho_i{p_o}^d=\phi_i/I_i$ for a pixel to be occupied by species $i$ is small.  In such cases the relative variance is more simply
\begin{equation}
	{\mathcal V}(L)\equiv \frac{ {\sigma_\phi}^2(L) }{\phi} = \frac{ \langle V_P \rangle }{L^d},
\label{vpixpoiss}
\end{equation}
where $\langle V_P\rangle = \sum \phi_i (I_i {p_o}^d) /\phi = \langle I_P \rangle {p_o}^d$ is the volume-fraction weighted average particle volume.

When the arrangement of particles is not random, i.e.\ if it has some degree of uniformity or order, the variance must be smaller than the upper bound given by Eqs.~(\ref{vpixmulti},\ref{vpixpoiss}).  Then we may define a hyperuniformity disorder length $h(L)$ such that fluctuations occur only for particles lying in the boundary volume $[L^d-(L-2h)^d]$ of thickness $h$ near the surface of the measuring windows \cite{ATCpixel}.  Specifically, $h$ is defined from ${\mathcal V}_{data}(L)=( {\sigma_\phi}^2/\phi )_{data}$ by the equivalent expressions
\begin{eqnarray}
	{\mathcal V}_{data}(L) &=& \frac{ \langle V_P \rangle }{L^d}\left[ \frac{L^d - (L-2h)^d }{L^d} \right], \label{vpixgen} \\
	                &=& {\mathcal V}(L)-{\mathcal V}(L-2h)\left( \frac{L-2h}{L} \right)^d, \label{vpixgenB}
\end{eqnarray}
where ${\mathcal V}(L)=\langle V_P\rangle/L^d$ is the relative variance for a random arrangement of pixel particles.  We shall see that the first of these expressions also holds for extended particles in the limit $L^d\gg \langle V_P\rangle$, and that the second holds in general where ${\mathcal V}(L)$ is the relative variance for a random arrangement of the same objects.


\section{Extended Particles}

Analysis of ${\mathcal V}_{data}(L)$ measurements is based on comparison with the prediction ${\mathcal V}(L)$ for a totally random arrangement of particles of the same type.  In this section we develop the necessary machinery and put it to use for a few different extended particles.

We begin with general monodisperse particles of volume $V_P$, number density $\rho$, and average volume fraction $\phi=\rho V_P$.  The average number of particles that overlap with measuring windows of volume $V_\Omega$ is $\overline N = \rho V_R$, where $V_R$ is the volume of a region that is larger than $V_\Omega$ according to the non-zero size of the particles.   This can be written as $\overline N = (V_R/{p_o}^d)q$ where $(V_R/{p_o}^d)$ is the number of pixels on which an overlapping particle may be centered, and $q=\rho {p_o}^d = \phi {p_o}^d/V_P$ is the probability for a given pixel to have a particle centered upon it.  For a random arrangement the variance in the number of overlapping particles is then either ${\sigma_N}^2 = {\overline N}$ if particles are placed totally at random (Poisson statistics), or  ${\sigma_N}^2 = {\overline N}(1-q)$ if particle centers are not allowed to overlap (binomial statistics).  If the image resolution is good, then $q$ is small and the distinction between Poisson versus binomial randomness vanishes.  Therefore we henceforth assume Poisson statistics, without much loss of generality.  Next, the volume fraction variance is given by the volume variance as ${\sigma_\phi}^2={\sigma_V}^2/{V_\Omega}^2$.  In turn the volume variance is ${\sigma_V}^2 = {\sigma_N}^2 \langle {V_Q}^2\rangle$ where $\langle {V_Q}^2\rangle$ is the mean-squared overlap volume for particles that are at least partially inside the measuring window.  Combining these ingredients, the relative variance defined by ${\mathcal V}(L)={\sigma_\phi}^2(L)/\phi$ for a random arrangement is
\begin{equation}
	\boxed{   {\mathcal V}(L) = \frac{V_R\langle{V_Q}^2\rangle }{V_P {V_\Omega}^2}.   }
\label{hudls1}
\end{equation}
This is the first fundamental equation of HUDLS.  Four different volumes are involved, all of which depend on the window size except for the particle volume.  Eq.~(\ref{hudls1}) holds for any shape of particle and measuring window, and for continuous or pixelated space of any dimension.  Note, crucially, that $\phi$ does not appear on the right-hand side; therefore, the volume fraction-dependence of ${\sigma_\phi}^2(L)$ for random patterns is exactly canceled by the normalization factor of $\phi$.   For random patterns, the relative variance is independent of $\phi$ and may be computed from the right-hand size of Eq.~(\ref{hudls1}) based on just the geometries of a single particle and the measuring window.

\subsection{Polydispersity}

Real systems are rarely monodisperse.  For the general case of polydisperse particles, the volume fraction $\phi=\sum\phi_i$ is the sum over different particle species with volume fractions $\phi_i=\rho_iV_{Pi}$ set by the individual number densities and particle volumes.  For random configurations with Poisson statistics, the above argument works through to give ${\mathcal V}=\sum W_i [V_{Ri}\langle V_{Qi}\rangle/(V_{Pi}{V_\Omega}^2)]$ where $W_i=\phi_i/\phi$.  Thus polydispersity is handled by a volume fraction-weighted average of the monodisperse expectation, just as seen earlier for pixel particles.  For the remainder of the theory section we thus focus attention on evaluating the right-hand side of Eq.~(\ref{hudls1}) for individual particles of various shape.

\subsection{Limits}

To check calculation results, the limiting behavior of Eq.~(\ref{hudls1}) for small and large measuring windows can be evaluated as follows.  For small $L$, and pixelated space, we first write the particle volume as $V_P=\sum I {p_o}^d=n_P\langle I_P\rangle {p_o}^d$, where $n_P$ is the number of pixels covered by the particle and $\langle I_P\rangle$ is the average intensity of the pixels in a particle.   The smallest measuring window is $V_\Omega={p_o}^d$, i.e.\ one voxel, for which the measuring window equals the volume $V_R=n_P{p_o}^d$ covered by all $n_P$ particle pixels.  The mean-squared overlap volume is therefore $\langle {V_Q}^2 \rangle = (1/n_P)\sum( I {p_o}^d)^2=\langle {I_P}^2\rangle {p_o}^{2d}$ where $\langle {I_P}^2\rangle$ is the mean-squared intensity of all the pixels in a particle.  Plugging into the right-hand size of Eq.~(\ref{hudls1}), the relative variance is thus expected to have an intercept of
\begin{equation}
	{\mathcal V}(p_o)=\langle {I_P}^2\rangle /\langle I_P\rangle.
\label{SsmallL}
\end{equation}
This reduces to $I_o$ for pixel particles of volume $V_P=I_o {p_o}^d$, as expected from Eq.~(\ref{vpixpoiss}).  Note that Eq.~(\ref{SsmallL}), and the analogous limit ${\mathcal V}(0)=\langle {I_P}^2\rangle /\langle I_P\rangle$ for continuous space, both hold even if the measuring windows are not step functions. 

For very large measuring windows, $V_\Omega\gg V_P$, partially-overlaping particles are far less numerous than fully-enclosed particles and hence $V_R=V_\Omega$ becomes a good approximation.  And similarly the mean-squared overlap becomes $\langle {V_Q}^2\rangle = \langle H^2\rangle{V_P}^2$ where $H({\bf x})$ is a hat function that specifies the measuring window, normalized such that $V_\Omega=\int H({\bf x})d{\bf x}$, $\langle H\rangle=1$, and $\langle H^2\rangle = \int H^2({\bf x})d{\bf x}/V_\Omega$.  For pixelated space, these integrals become discrete sums.  With these ingredients, the relative variance for random patterns is then expected from Eq.~(\ref{hudls1}) to vanish as
\begin{equation}
	{\mathcal V}(L)\rightarrow \langle H^2\rangle V_P/V_\Omega.
\label{SlargeVB}
\end{equation}
For step-function measuring windows of volume $V_\Omega=L^d$, and any shape, $\langle H^2\rangle=1$ holds and the limiting behavior becomes
\begin{equation}
	{\mathcal V}(L)\rightarrow  V_P/L^d.
\label{SlargeL}
\end{equation}
This agrees with the known result that the variance decays as $1/L^d$ for random point patterns, and shows how the proportionality constant exactly equals the particle volume.  It also matches Eq.~(\ref{vpixpoiss}); therefore, for large $L$, the relative variance for extended particles becomes equal to that for the central pixel representation.

\subsection{Computation Methods}

In evaluating Eq.~(\ref{hudls1}) for the given shapes of the particle and measuring window, only the numerator poses difficulty.  For continuum particles and windows, one approach is to use a Fourier method since the overlap of a particle $I({\bf x})$ with a measuring window $H({\bf x})$ can be written as a convolution.  The variance is set by the square of this overlap, averaged over possible relative placements of particle and window.  This is given by Parseval's theorem as
\begin{equation}
	V_R\langle {V_Q}^2\rangle = \int | \tilde H \tilde I |^2   d{\bf k}/(2\pi)^d.
\label{parseval}
\end{equation}
For a cubic measuring window of volume $V_\Omega=L^d$, for example, the transform of the boxcar hat function is $\tilde H({\bf k})=L^d {\rm sinc}(k_xL/2){\rm sinc}(k_yL/2)\cdots$.

The numerator of Eq.~(\ref{hudls1}) may also be evaluated by direct integration in the continuum limit.  For clarity and for ease of translating to discrete sums for pixelated images, we write it out explicitly in one dimension.  The particle is imagined to extend from $x=0$ to $x=p_m$ and to have ``volume'' $V_P=\int_0^{p_m}I(x)dx$.  Two nontrivial kinds of overlap are involved:
\begin{eqnarray}
	O_1(x_o) &=& \int_0^{x_o} I(x) dx, \label{O1continuum} \\
	O_2(x_o) &=& \int_{x_o}^{x_o+L}I(x)dx \label{O2continuum}
\end{eqnarray}
The first is for when the particle extends only some distance $x_o$ into the measuring window;  the second is for when the entire window is covered by a portion of the particle.  The third kind of overlap is a constant, $O_3=V_P$, for when the particle is entirely inside a step-function measuring window.  The numerator of Eq.~(\ref{hudls1}) is given by squaring these and integrating over all possible relative placements $x_o$ of particle and window:
\begin{widetext}
\begin{numcases}{V_R\langle {V_Q}^2 \rangle = }
	\label{VRVQ2continuumA}
		 2\int_0^L [O_1(x_o)]^2 dx_o + \int_0^{p_m-L}[O_2(x_o)]^2 dx_o & $L\le p_m$, \\
	\label{VRVQ2continuumB}
		 2\int_0^{p_m}[O_1(x_o)]^2 dx_o + {V_P}^2(L-p_m) & $L\ge p_m$.
\end{numcases}
\end{widetext}
The factors of $2$ appear because the particles are assumed to be symmetric and can extend part-way into the window from either side.

For one dimensional pixelated images, we imagine the particles to cover pixels $i=1$ to $i=p_m/p_o$ and to have ``volume'' $V_P = \sum_{i=1}^{p_m/p_o}I(i)p_o$.  As above there are two non-trivial particle-window overlap possibilities,
\begin{eqnarray}
	O_1(n) &=& \sum_{i=1}^n I(i)p_o, \label{O1pixelated} \\
	O_2(n) &=& \sum_{i=n}^{n+\frac{L}{p_o}-1} I(i)p_o.  \label{O2pixelated}
\end{eqnarray}
The first is for when the first $n$ pixels of the particle extend into the measuring window;  second is for when the entire window is covered by a portion of the particle starting at pixel $n$.  And similar to the continuum case, the numerator of Eq.~(\ref{hudls1}) is given by squaring these and summing over all possible relative placements $n$ of particle and window:
\begin{widetext}
\begin{numcases}{V_R\langle {V_Q}^2 \rangle = }
	\label{VRVQ2pixelatedA}
		 2\sum_{n=1}^{\frac{L}{p_o}-1} [O_1(n)]^2 p_o + \sum_{n=1}^{\frac{p_m-L}{p_o}+1}[O_2(n)]^2 p_o & $L\le p_m$, \\
	\label{VRVQ2pixelatedB}
		 2\sum_{n=1}^{\frac{p_m}{p_o}-1}[O_1(x_o)]^2 p_o + {V_P}^2(L-p_m+p_o) & $L \ge p_m$.
\end{numcases}
\end{widetext}
Here the final factor $(L-p_m+p_o)$ comes from $p_o$ times the number of ways to place the particle entirely inside the window.


\subsection{Results}

The first several examples are for cubic step-function measuring windows, $V_\Omega=L^d$, for both pixelated and continuous space, and a variety of different particles.  This is followed by examples with radially-symmetric spherical and Gaussian measuring windows, for continuous two- and three-dimensional space.  These should cover most cases of interest for analyzing experiments and simulations.

\subsubsection{Pixel particles}

The easiest use of Eq.~(\ref{hudls1}) is for the case of monodisperse pixel particles of volume $V_P=I_o{p_o}^d$.  Since such particles either lie entirely inside or entirely outside the measuring windows, the measuring region and window are equal, $V_R=V_\Omega=L^d$.  And the mean-squared overlap is exactly $\langle {V_Q}^2\rangle={V_P}^2$.  Plugging these four volumes into Eq.~(\ref{hudls1}) then gives the relative variance for a random arrangement of pixel particles as
\begin{equation}
	{\mathcal V}(L)=V_P/V_\Omega=I_o{p_o}^d/L^d.
\label{Spix}
\end{equation}
This is exact, for any packing fraction and any window size, and recovers the prior result quoted in Eq.~(\ref{vpixpoiss}).

\subsubsection{Rectangular particles}

The first new case is for particles with constant intensity $I_o$ that cover a rectangular region of more than just one pixel.  We start with a one dimensional rectangular particle of length $p$, that covers $p/p_o$ pixels and has volume $V_P=I_o p$.  The sums in Eqs.~(\ref{O1pixelated}-\ref{VRVQ2pixelatedB}) with $I(i)=I_o$ are readily evaluated.  Dividing $V_R\langle {V_Q}^2 \rangle$ by $V_P{V_\Omega}^2 =I_o p L^2$ then gives the predicted relative variance for random particle placements as
\begin{numcases}{{\mathcal V}_1(L,p) = I_o}
	\label{S1rectA}
		  \frac{L p - (L^2 - p_o^2)/3}{Lp} & $L\le p$, \\
	\label{S1rectB}
		 \frac{L p - (p^2 - p_o^2)/3}{L^2} & $L \ge p$.
\end{numcases}
This function satisfies four checks:  It is continuous at $L=p$, satisfies the expected limits ${\mathcal V}_1(p_o)=I_o$ and ${\mathcal V}_1(L)\rightarrow I_o(p/L)$ for large $L\gg p$, and for $p=p_o$ reduces to the $d=1$ pixel particle result of Eq.~(\ref{Spix}).  The limit of continuous space, where there is a continuum of measuring window overlap possibilities, is given by taking $p_o=0$.  The form of the variance in this limit may be verified two ways.  The first is by the Fourier method, Eq.~(\ref{parseval}), using a particle transform of $\tilde I(k)=I_op{\rm sinc}(kp/2)$.  The second is by direct integration using Eqs.~(\ref{O1continuum}-\ref{VRVQ2continuumB}).  Note that the effect of pixelated space (i.e.\ of $p_o>0$) is non-trivial in that the form of Eqs.~(\ref{S1rectA}-\ref{S1rectB}) cannot be guessed from the continuum limit, e.g.\ by supposing $L\rightarrow L-p_o$.

For a rectangular particle $V_P = I_o p_x p_y\cdots$ in higher dimensions, with cubic measuring windows, the integrals and sums are all separable.  Therefore the relative variance is the product
\begin{equation}
	{\mathcal V}(L)={\mathcal V}_1(L,p_x){\mathcal V}_1(L,p_y)\cdots
\label{Sd}
\end{equation}
of the 1-dimensional result (with just one factor of $I_o$).  For small $L$, the leading behavior is ${\mathcal V}(L)/I_o = 1 - 2(\sum1/p_i)(L-p_o)/3 +\mathcal O(L-p_o)^2$.  But if $p_o$ is first set to zero, then the expansion is ${\mathcal V}(L)/I_o = 1 -( \sum1/p_i)(L/3) + \ldots$ with no factor of two.  For large $L$, the asymptotic behavior is $\mathcal V(L)=V_P/L^d$.

\subsubsection{Continuum examples}

In experimental grayscale images, the intensity profile is typically brightest in the middle of the particle.  For continuum Gaussian particles in $d=1$ dimensions, we take $I(x) = I_o\sqrt{2/\pi}\exp[-2(x/p)^2]$.  The corresponding volume is $V_P=I_o p$ where the ``particle length'' $p$ is twice the standard deviation of the intensity profile.  Using the Fourier method, we find the relative variance to be
\begin{equation}
	{\mathcal V}_1(L) = I_o \frac{ \sqrt{\pi}(L/p){\rm erf}(L/p)-\{1-\exp[-(L/p)^2]\}}{ \sqrt{\pi}(L/p)^2 }.
\label{Sgaussian}
\end{equation}
The intercept is ${\mathcal V}_1(0)=I_o/\sqrt{\pi}$, in accord with $\langle {I_P}^2\rangle/\langle I_P\rangle$.  For large $L$ the relative variance expands as ${\mathcal V}_1(L)=I_o[ (p/L)-(p/L)^2/\sqrt{\pi}+\mathcal{O}(1/L^3)]$, with the expected leading behavior.  For higher dimensions, just as for rectangular particles, the $d=1$ result may be multiplied together according to Eq.~(\ref{Sd}).  We were unable to compute the relative variance for pixelated Gaussian particles.  Judging from the pixelated result for rectangular particles, it would be quite different from substituting $L\rightarrow L-p_o$ in Eq.~(\ref{Sgaussian}).  For real images, which are pixelated, Eq.~(\ref{Sgaussian}) becomes correct in the limit $p\gg p_o$ that the particles are large.

As another one dimensional continuum example, we take $I(x)=I_o (2/\pi)/[1+(2x/p)^2]$.  This is a Lorentzian particle with volume $V_P=I_o p$, where $p$ is the full-width half-max of the profile.  Using the Fourier method we compute the relative variance as
\begin{equation}
	{\mathcal V}_1(L)=I_o \frac{2(L/p)\arctan(L/p)-\ln[1+(L/p)^2]}{\pi (L/p)^2}.
\label{Slorentz}
\end{equation}
The intercept is ${\mathcal V}_1(0)=I_o/\pi$ and the large-$L$ behavior is ${\mathcal V}_1(L)=I_o[(p/L)-(2/\pi)(p/L)^2+\mathcal{O}(1/L^4)]$, both in agreement with the expected limits.  For higher dimensions, the product of Eq.~(\ref{Sd}) does not correspond to a radially-symmetric Lorenztian particle since the position variables in the intensity profile are not separable.

\subsubsection{Sine-squared particles}

Since it was not possible to compute a pixelated version of a Gaussian particle, we tried a few approximate alternatives.  We succeeded with parabolic and quartic profiles of form $I(i)=I_{max}\{4(i-1/2)(p_o/p_m)[1-(i-1/2)(p_o/p_m)]\}^m$ with $1\le i\le p_m/p_o$; however, the results are quite messy even for the particle volume.  Perhaps surprisingly, the sums can be evaluated and are actually simpler for sine-squared particles: $I(i)=I_o \sin^2[\pi(i-1/2)p_o/p]$, with $1\le i \le 2p/p_o$ and $p$ being the full-width half-max.  For this profile, the volume is $V_P=I_o p$, exactly, and the relative variance is
\begin{widetext}
\begin{numcases}{{\mathcal V}_1(L) = I_o}
	\label{S1sine2A}
		\frac{ 2L(L^2-6Lp-p_o^2)(C_1-1)+3p_o^2[2p+(L-2p)C_L+4LC_2] - 3p_o^3(C_1/S_1+2/S_2)S_L  }{48 L^2 p {S_2}^2} & $L\le 2p,$~~~~~~  \\
	\label{S1sine2B}
		 \frac{ (6Lp - 4p^2+p_o^2) +(3p^2/\pi^2)(1+4C_2)/{S_c}^2         }{6L^2} & $L \ge 2p,$
\end{numcases}
\end{widetext}
where $C_1=\cos(\pi p_o/p)$, $C_2=\cos[\pi p_o/(2p)]$, $C_L=\cos(\pi L/p)$,  $S_1=\sin(\pi p_o/p)$, $S_2=\sin[\pi p_o/(2p)]$, $S_L=\sin(\pi L/p)$, and $S_c={\rm sinc}[\pi p_o/(2p)]$.  This result is verified to be continuous at $L=2p$ and to have the expected limits of ${\mathcal V}_1(p_o)=\langle {I_P}^2\rangle/\langle I_P\rangle=(3/4)I_o$ and ${\mathcal V}_1(L)=I_o[(p/L)+\mathcal{O}(p/L)^2]$.  As a further check, the continuum limit of $p_o\rightarrow 0$ matches the result from the integration method, Eqs.~(\ref{O1continuum}-\ref{VRVQ2continuumB}).

\subsubsection{Spherical measuring windows}

Prior work typically uses spherical measuring windows, since this is simple to implement with the randomly-placed measuring window method for computing the variance.  It's not obvious how do this for pixelated space.  In the continuum limit, the Fourier transform of a radial unit step function is $\tilde H_R(k)=2\pi R^2 J_1(kR)/(kR)$ and $\tilde H_R(k)=4\pi R^3[ \sin(kR)-kR\cos(kR)]/(kR)^3$ in two and three dimensions, respectively, where $k=|{\bf k}|$ and $J_1(x)$ is the Bessel function of the first kind.  For $d=2$ circular particles of radius $r$ and intensity $I_o$, and circular measuring windows of radius $R$, the relative variance is then given by the Fourier method as
\begin{equation}
	{\mathcal V}_2(R)=8I_o\int_0^\infty \left[ \frac{ J_1(x)}{x} \frac{J_1(x R/r)}{x R/r}\right]^2 xdx
\label{circs}
\end{equation}
where $x=kr$ is a dimensionless integration variable.  This evaluates to a large number of algebraic, logarithmic, and polylogarithmic terms, as well as separate cases for when the measuring window is larger or smaller than the particles.  
For $d=3$, the corresponding prediction for random arrangement of constant-intensity spherical particles of radius $r$ is much simpler:
\begin{numcases}{{\mathcal V}_3(R) = I_o}
	\label{Vss1}
		 1-\frac{27R}{35r}+\frac{2R^3}{21r^3} & $R\le r$, \\
	\label{Vss2}
		 \frac{r^3}{R^3}-\frac{27r^4}{35R^4}+\frac{2r^6}{21R^6}& $R\ge r$.
\end{numcases}
This has the correct limits and is continuous at $R=r$.

The difference between these results and Eqs.~(\ref{S1rectA}-\ref{Sd}) for cubic windows is maximal at about 3\% and 10\% in two and three dimensions, respectively, when particles and measuring windows are nearly same size.  On this basis, compact step-function particles and measuring windows in general could be roughly analyzed using the rectangular particles / cubic windows prediction of Eqs.~(\ref{S1rectA}-\ref{Sd}).  Even more roughly and simply, the relative variance function could be approximated by the rational function
\begin{equation}
	{\mathcal V}_d(x) = I_o\left[ \frac{x+3}{x^2 + 2x +3} \right]^d,
\label{Vrat}
\end{equation}
where $x$ is the ratio of window to particle width.  This matches the initial and final decays, as well as the value at $L=p$, for cubic particles with cubic measuring windows, in all dimensions.

For Gaussian particles with intensity profile $I({\bf x}) = I_o(2/\pi)^{d/2}\exp[-2({\bf x}/r)^2]$ and volume $V_P=I_o r^d$, the second intensity moment is $\langle {I_P}^2\rangle/\langle I_P\rangle=I_o/\pi^{d/2}$, and relative variances for circular and spherical measuring windows of radius $R$ in two and three dimensions are respectively found by the Fourier method to be
\begin{widetext}
\begin{eqnarray}
	{\mathcal V}_2(R) &=& I_o\frac{1-\left\{I_1[2(R/r)^2]+H_G[(R/r)^4]   \right\}\exp[-2(R/r)^2] }{\pi (R/r)^2}, \label{Vgc} \\
	{\mathcal V}_3(R) &=& I_o\frac{ 1-6(R/r)^2-[1-2(R/r)^2]\exp[-4(R/r)^2]+4\sqrt{\pi}(R/r)^3 {\rm erf}(2R/r) }{(16\pi^{3/2}/3)(R/r)^3 }, \label{Vgs}
\end{eqnarray}
\end{widetext}
where $I_1(x)$ is the modified Bessel function of the first kind and $H_{G}(x)$ is the confluent hypergeometric function $_0F_1(;1;x)$, given by {\tt Hypergeometric0F1Regularized[1,x]} in Mathematica for example.  The limits all behave correctly as ${\mathcal V}_2(R)=(I_o/\pi)[1-(R/r)^2+\mathcal{O}(R/r)^4]$ and ${\mathcal V}_2(L)=(I_o/\pi)[(r/R)^2-(r/R)^3/\sqrt{\pi}+\mathcal{O}(r/R)^5]$ in two dimensions, and ${\mathcal V}_3(R)=(I_o/\pi^{3/2})[1-(6/5)(R/r)^2+{\mathcal O}(R/r)^4]$ and ${\mathcal V}_3(R)=[3I_o/(4\pi)][(r/R)^3-(3/2)(r/R)^4/\sqrt{\pi}+\mathcal{O}(r/R)^5]$ in three dimensions.

\subsubsection{Gaussian measuring windows}

Measuring windows need not be step functions.  As a potentially useful continuum example, we consider a radial Gaussian measuring window $H({\bf x})=(2/\pi)^{d/2}\exp[-2({\bf x}/L)^2]$ where $L$ is twice the standard deviation in each dimension.  These may be helpful for smoothing over short-range noise or features, and could be realized optically.  For this hat function, the volume is $V_\Omega=L^d$, the mean-squared value is $\langle H^2\rangle = 1/\pi^{d/2}$, and the Fourier transform is $\tilde H({\bf k})=L^d\exp[-({\bf k}L)^2/8]$.  For asymmetric Gaussian particles $I(x,y,\ldots)=I_o(2/\pi)^{d/2}\exp[-2(x/p_x)^2-2(y/p_y)^2-\ldots]$, with volume $V_P=I_o p_x p_y\cdots$, the relative variance is found from both direct integration and the Fourier method to be
\begin{equation}
	{\mathcal V}(L) = \frac{ I_o p_x p_y\cdots}{\sqrt{\pi^d(L^2 + p_x^2)(L^2 + p_y^2)\cdots}}.
\label{gg}
\end{equation}
This result applies for any orientation of the particles, since the measuring windows are radially symmetric.  Note that ${\mathcal V}(0)=I_o/\pi^{d/2}$ is the same intercept as for Gaussian particles with a square measuring window, as expected.  And the large-$L$ limit is $V_P/(\pi^{d/2}L^d)$, in accord with Eq.~(\ref{SlargeVB}).

For solid circular and spherical particles of radius $r$ and intensity $I_o$ in two and three dimensions, with a Gaussian measuring window, the respective relative variances may also be found by the Fourier method:
\begin{widetext}
\begin{eqnarray}
	{\mathcal V}_2(L) &=& I_o- I_o\left\{ I_1[2(r/L)^2] + H_{G}[(r/L)^4] \right\}\exp[-2(r/L)^2], \label{Vcg} \\
	{\mathcal V}_3(L) &=& I_o\frac{ 1-6(r/L)^2-[1-2(r/L)^2]\exp[-4(r/L)^2]+4\sqrt{\pi}(r/L)^3 {\rm erf}(2r/L) }{4\sqrt{\pi}(r/L)^3 }. \label{Vsg}
\end{eqnarray}
\end{widetext}
Apart from an overall factor, these are identical to the earlier Gaussian-particle / spherical-window expressions, where $R/r$ is replaced by $r/L$; this particle-window duality is expected from the symmetry of particle/window convolution in Eq.~(\ref{parseval}).  Here, the limiting behaviors all check out correctly as ${\mathcal V}_2(L)=I_o[1-L/(\sqrt{\pi}r)+\mathcal{O}(L/r)^3]$ and ${\mathcal V}_2(L)=I_o[(r/L)^2-(r/L)^4+\mathcal{O}(r/L)^6]$ in two dimensions, and ${\mathcal V}_3(L)=I_o[1-3L/(2\sqrt{\pi}r)+{\mathcal O}(L/r)^3]$ and ${\mathcal V}_3(L)=I_o[4r^3/(3\sqrt{\pi}L^3)-8r^5/(5\sqrt{\pi}L^5)+\mathcal{O}(r/L)^5]$ in three dimensions.


\section{Validation}

\begin{figure}[ht]
\includegraphics[width=3.000in]{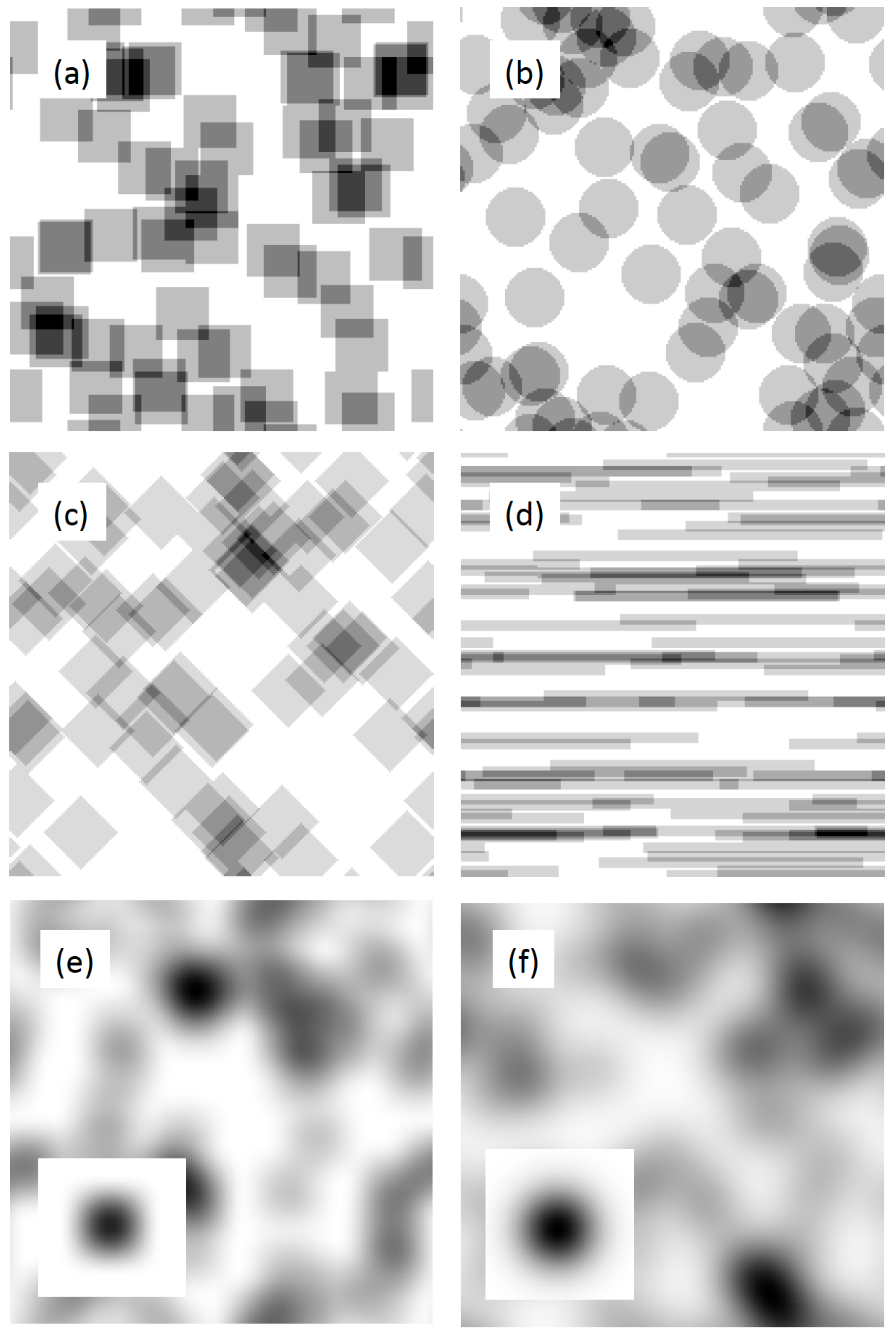}
\caption{Poisson arrangement of various extended particles: (a) $25\times25$ square, (b) circular, (c) diamond, (d) $5\times125$ rectangular, (e) sine-squared, (f) Gaussian.  For each the image size is $200\times200$ square pixels, the total area fraction is $\phi=1$, and the particle area is approximately $25\times25$ square pixels (exactly for square, rectangular, and sine-squared particles).  The insets in (e,f) show isolated particles.}
\label{Particles}
\end{figure}

To test the above methodology and some of the specific predictions, we now create and analyze two-dimensional random arrangements of particles of various shapes and packing fractions.  Small example patterns are shown in Fig.~\ref{Particles} for six different particle shapes, all with the same particle volume $V_P=(25p_o)^2$ and packing fraction $\phi=1$, and with periodic boundary conditions (to be used throughout).  These are Poisson patterns, where the particle locations are chosen totally at random using a random-number generator.  At high packing fractions, as shown, there is considerable particle-particle overlap.  Close inspection reveals pixelation effects in both particle placement and shape, even though the particles are fairly large compared to the pixel width.  Note also that the sine-squared particles look roughly Gaussian, but are not radially symmetric and are not as extended.

\subsection{Rectangular Particles}

\begin{figure}[ht]
\includegraphics[width=3.000in]{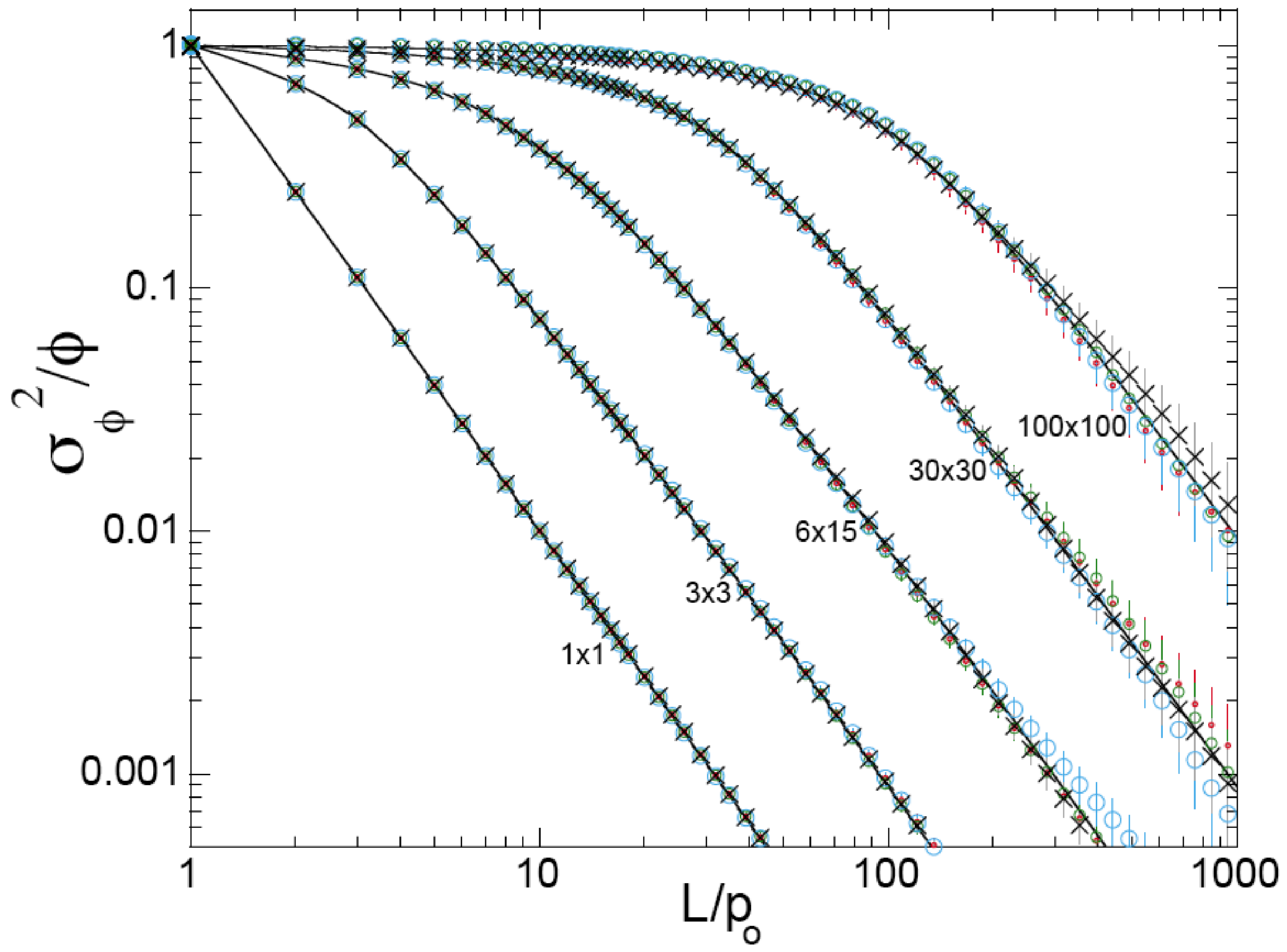}
\caption{(color online) Relative variance vs measuring window size for simulated Poisson patterns of rectangular particles, with dimensions as labeled in pixel units $p_o$.  For each particle width there are four volume fractions, $\phi=\{0.02,~0.15,~1.0,~5\}$, indicated by increasing symbol size except for $\times$ for $\phi=5$ (500\%).  To within statistical uncertainty these collapse together and agree with the prediction of Eqs.~(\ref{S1rectA}-\ref{Sd}), shown by the solid curves.  The simulation system size is $3000\times3000$ square pixels, and has periodic boundary conditions.}
\label{RectPoiss}
\end{figure}

As the first test, we illustrate behavior versus particle width and packing fraction for rectangular particles.  We choose five different particle widths, ranging from $1\times1$ up to $100\times100$ square pixels, and four different area fractions, $\phi=\{0.02,~0.15,~1,~5\}$.  For each combination we create Poisson patterns of size $(3000p_o)^2$, and compute the relative variance ${\mathcal V}_{data}(L)={\sigma_\phi}^2(L)/\phi$ using the Fourier method described in the Image Processing section.  The window sizes $L$ are chosen on a logarithmic scale from $L=p_o$ to $L=1500p_o$.  For larger $L$ beyond this range, finite-size effects cause a strong systematic decrease in the measured variance \cite{DreyfusPRE2015}.  Final simulation results for the relative variance are plotted in Fig.~\ref{RectPoiss}.  As expected, dividing the variance by $\phi$ causes the data to collapse -- such that the relative variance is independent of $\phi$.  And for each particle width, the collapsed data closely matches the plotted predictions of Eqs.~(\ref{S1rectA}-\ref{Sd}).  In particular, the relative variance begins at ${\mathcal V}_{data}(p_o)=\langle {I_P}^2\rangle/\langle I_P\rangle=1$ and has a final asymptotic decay of ${\mathcal V}_{data}(L) \rightarrow V_P/L^2$.  The crossover to final scaling is set by particle width.

The error bars plotted in Fig.~\ref{RectPoiss} are given by Eq.~(\ref{DeltaSigma}) as $\Delta{\sigma_\phi}^2={\sigma_\phi}^2\sqrt{2/(s-1)}$ where the number of independent samplings is $s = ({\rm image\ area})/L^2$.  Note that the error bars therefore bloom with increasing $L$, because there are fewer independent samplings of an image by all the possible windows.  These error bars represent the statistical scatter expected for an ensemble of different images simulated under the same conditions.  They {\it do not} represent the statistical scatter between successive $L$-values for the spectrum of given image; indeed, with the Fourier method, ${\sigma_\phi}^2(L)$ versus $L$ is perfectly smooth.  While each data set in Fig.~\ref{RectPoiss} may exhibit an apparent smooth systematic deviation above or below the prediction, the different runs are seen to be scattered randomly around the prediction by an amount that is in accord with the plotted error bars.  Thus we conclude that the simulation results for both the relative variance {\it and} the uncertainty are in full agreement with prediction.

\subsection{Other Particle Shapes}

\begin{figure}[ht]
\includegraphics[width=3.000in]{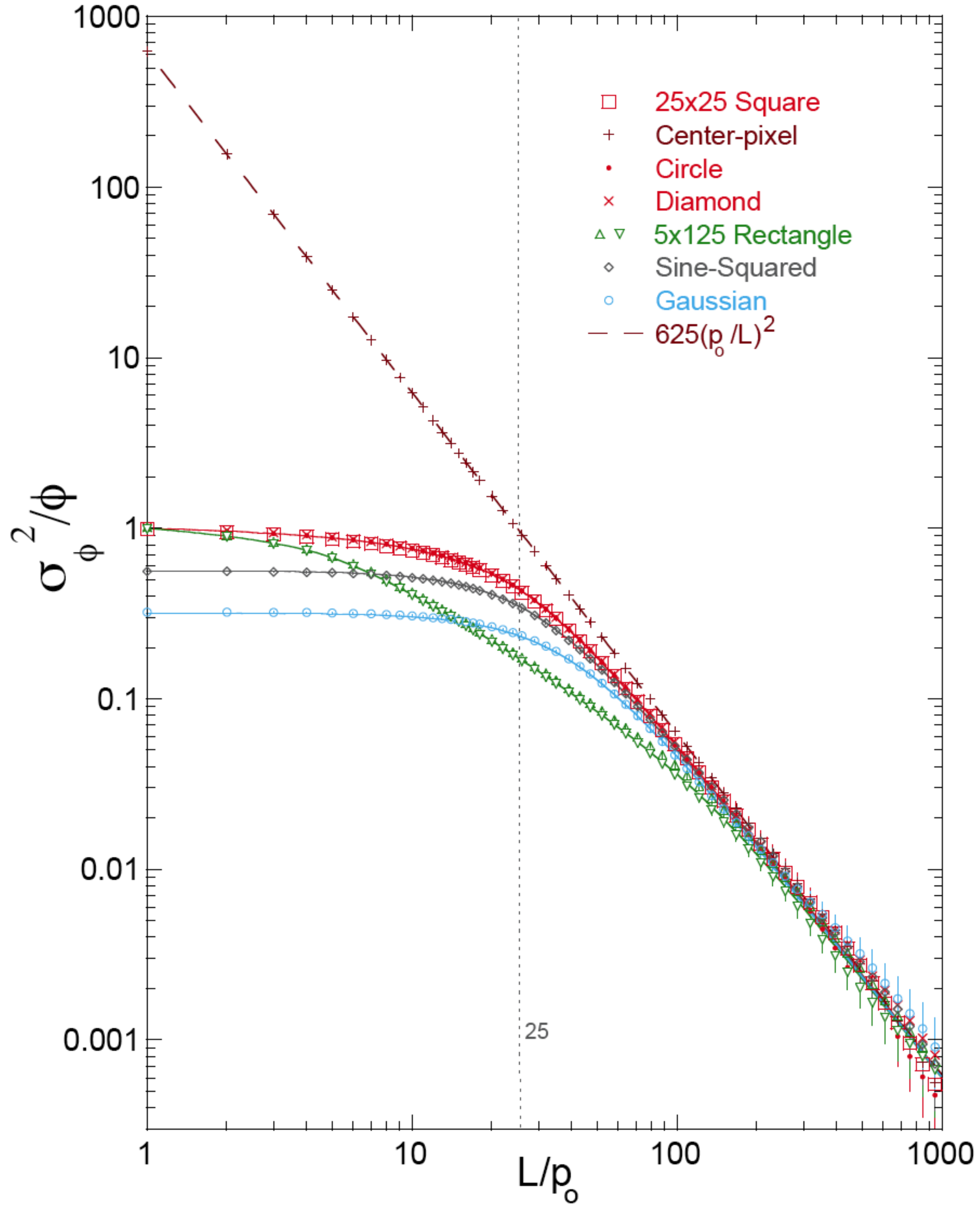}
\caption{(color online)  Relative variance vs measuring window size for simulated Poisson patterns of various particles, as labeled.  As per Fig.~\ref{Particles} the particle areas are approximately $V_P=(25p_o)^2$ (exactly for square, rectangular, and sine-squared cases), the total volume fraction is $\phi=1$, and the simulation system size is $3000\times3000$ square pixels.   The corresponding solid curves represent the predictions of Eqs.~(\ref{S1rectA}-\ref{Sd}) for rectangular particles, Eq.~(\ref{S1sine2A}-\ref{S1sine2B}) for sine-squared particles, and Eq.~(\ref{Sgaussian}) for Gaussian particles.  There are two data sets for the rectangular particles, one with all horizontal alignment and one with a 50:50 mixture of horizontal and vertical.  The $+$ symbols represent the relative variance results for the ``center-pixel'' pattern associated with the squares, where the entire weight of the particle is given to the central pixel.}
\label{Shapes}
\end{figure}

Next we consider the effect of particle shape, by simulating Poisson patterns made from the same six particles shown in Fig.~\ref{Particles}.   The particle volumes are made as close to $V_P=(25p_o)^2$ as possible.  This is achieved exactly for the square, rectangular, and sine-squared shapes, even though the latter is gray-scale.  The other areas are exactly $616{p_o}^2$ for pixelated-circular and $613{p_o}^2$ for pixelated-diamond, and approximately $622{p_o}^2$ for Gaussian.   For images of size $(3000p_o)^2$ with $\phi=1$ and periodic boundary conditions, relative variance results are collected in Fig.~\ref{Shapes}.  The corresponding predictions from the theory section are overlaid, and found to match the simulation data to within statistical uncertainty.  Note how the intercept ${\mathcal V}_{data}(p_o)$ is below 1 for the sine-squared and Gaussian grayscale particles, in agreement with the expected limit $\langle {I_P}^2\rangle / \langle I_P\rangle$.  And note how the final decay is the same for all six different particle shapes, ${\mathcal V}_{data}(L)\rightarrow V_P/V_\Omega=(25p_o/L)^2$.  Furthermore, the form of the crossover between limiting behaviors is seen to depend on particle size and shape.  This is particularly evident for the $5\times125p_o^2$ rectangular particles.  It is also interesting that the crossover is indistinguishable on this plot for the three compact binary particles: the square, the circle, and the diamond.

For contrast we also include in Fig.~\ref{Shapes} variance results for the central-pixel pattern associated with the squares.  For this, the center pixel of each square is set to a grayscale level of 625 and all other pixels in the square are set to zero.  Thus the entire weight of each particle is concentrated into one pixel, and $V_P=(25p_o)^2$ and $\phi=1$ still hold.  As seen, the variance agrees with the power-law expectation ${\mathcal V}(L)=V_P/L^2=625(p_o/L)^2$ for all $L$.

\subsection{Pixelation Versus Continuum}

\begin{figure}[ht]
\includegraphics[width=3.000in]{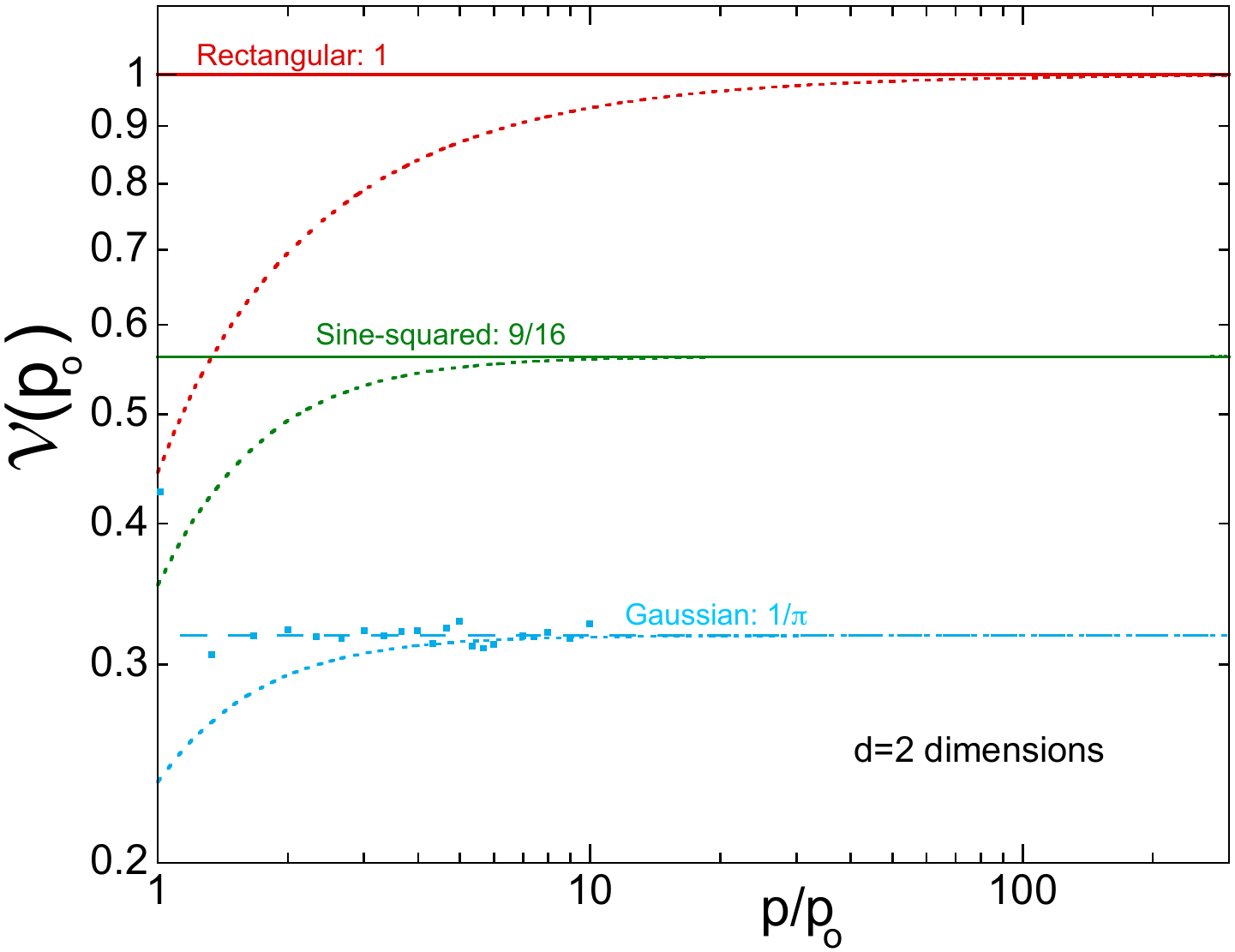}
\caption{(color online)  Relative variance at $L=p_o$ vs particle width $p$ for various $d=2$ dimensional particles, as labeled.  For pixelated particles this represents the intercept, and is shown by horizontal lines for square and sine-squared particles, and by symbols for simulated pixelated Gaussian particles.  For continuum particles, $\mathcal V(p_o)$ values are given by Eqs.~(\ref{S1rectA},\ref{Sd},\ref{S1sine2A}) with pixel width set to zero and by Eq.~(\ref{Sgaussian},\ref{Sd}) as-is; they are plotted as dashed curves, and are smaller than the $L=0$ intercept when the particle is not large.  The difference is a measure of how well pixelated particles may be approximated by continuum predictions.}
\label{Intercepts}
\end{figure}

The accuracy with which continuum predictions describe pixelated particles and measuring window locations may be studied by comparing the respective relative variances at $L=p_o$, where the difference is largest.  In particular, the relative variance intercepts are $\mathcal V(p_o)=I_o\langle {I_P}^2\rangle/\langle I_P\rangle$ at $L=p_o$ for pixelated particles, and $\mathcal V(0)=I_o\langle {I_P}^2\rangle/\langle I_P\rangle$ at $L=0$ for continuum particles;  therefore the value of $\mathcal V(p_o)$ for continuum particles must be lower.  For normalized $d=2$ dimensional particles, the intercepts are 1 for all binary particles and $(3/4)^d=9/16$ for the sine-squared particles, based on the above results -- in both pixelated and continuum limits.  For $d=2$ continuum Gaussian particles, the $L=0$ intercept is $1/\pi^d=1/\pi^2$; for pixelated Gaussian particles, the $L=p_o$ intercept may depend on particle size and is found by simulation.  Results for the pixelated $L=p_o$ intercepts are plotted versus particle width in Fig.~\ref{Intercepts} as solid horizontal lines for the square and sine-squared particles, and as symbols for the simulated pixelated Gaussian particles.  The latter is surprisingly constant.  For comparison, the continuum limits of the relative variance functions are evaluated $L=p_o$ and plotted versus $p$ using dashed curves.  These all start low, and rise up to the expected constant at large $p$.  The difference shows that the continuum approximation of pixelated particles at $L=p_o$ becomes better than one percent for square particles of size $p>60p_o$, for sine-squared particles of size $p>7p_o$, and for Gaussian particles of size $p>6p_o$.  For $p>2p_o$ pixelated Gaussian particles, the continuum approximation is still quite good -- better than eight percent.  For larger $L>p_o$, the continuum approximations must be even better.  Pixelation effects for particle and window locations are largest for binary particles and smallest for Gaussian particles.

\subsection{Polydispersity}

\begin{figure}[ht]
\includegraphics[width=3.000in]{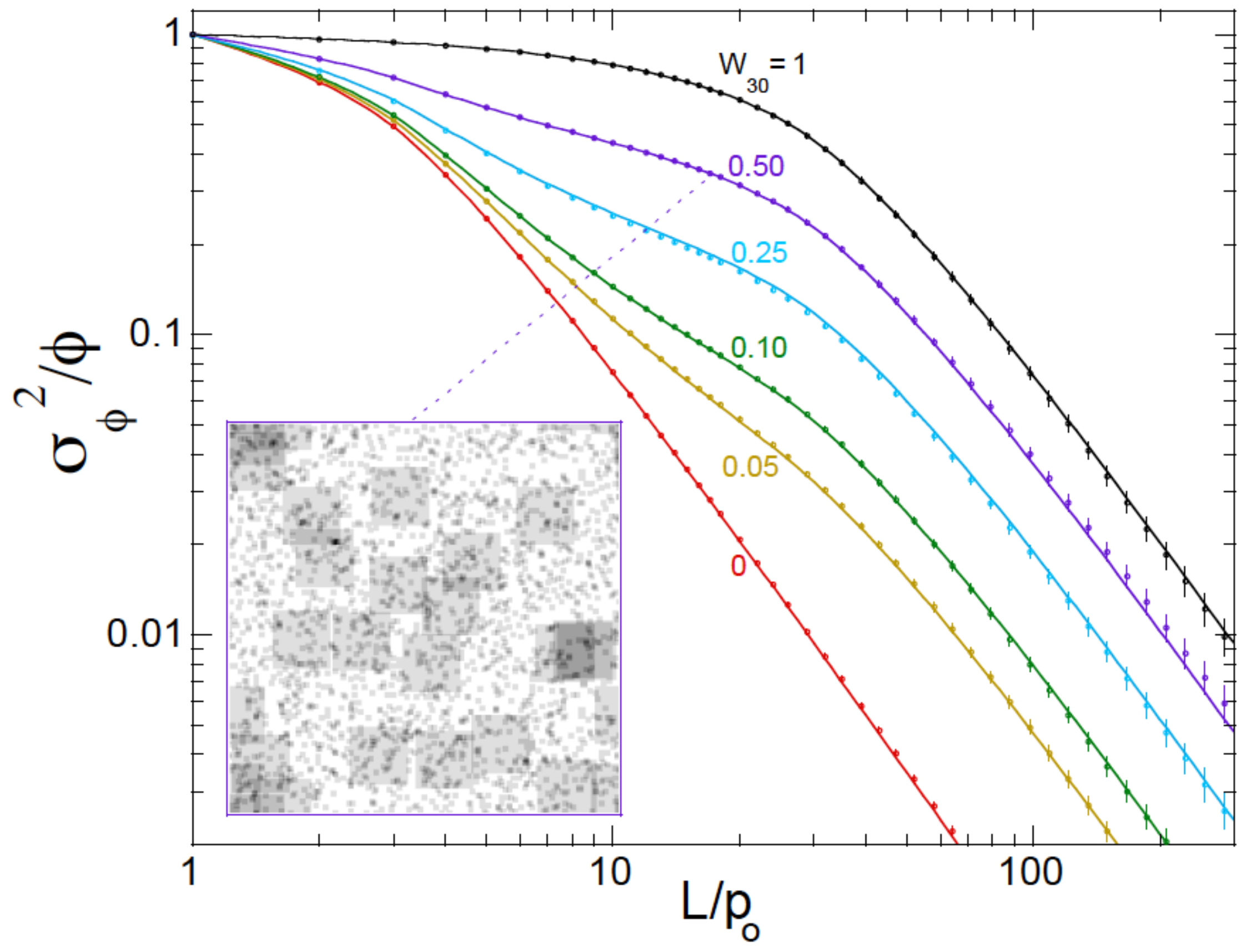}
\caption{(color online)  Relative variance vs measuring window size for simulated Poisson patterns of bidisperse mixtures of $3\times3$ and $30\times30$ square particles, labeled by the area-fraction weight $W_{30}$ of the larger particles.  The total volume fraction is $\phi=1$, and the simulation system size is $3000\times3000$ square pixels.   The corresponding solid curves represent the predictions of Eqs.~(\ref{S1rectA}-\ref{Sd}), computed separately for each particle size and averaged together with area-fraction weighting.  The inset shows a $200\times200$ sample with equal area fractions for the two species.}
\label{Poly}
\end{figure}

In the theory section we argued that polydispersity is accounted for by a volume-fraction weighted average over the different particle species.  This is tested by simulation data for polydisperse mixtures of $(3p_o)^2$ and $(30p_o)^2$ square particles in Fig.~\ref{Poly}.  As in prior simulations, the image sizes are $(3000p_o)^2$ with periodic boundary conditions, and the total volume fraction is $\phi=\phi_{3}+\phi_{30}=1$.  The weights are $W_3 = \phi_3/(\phi_3+\phi_{30})$ and $W_{30} = \phi_{30}/(\phi_3+\phi_{30}) = 1-W_3$, where the individual volume fractions equal the product of particle area and number density.  We simulate six different mixtures, with $W_{30}=\{0, 0.05, 0.10, 0.25, 0.50, 1\}$.  The relative variance results are plotted in Fig.~\ref{Poly}, along with the expectation based on the $W$-weighted averages of Eqs.~(\ref{S1rectA}-\ref{Sd}).  To within statistical uncertainty, there is perfect agreement.  It is worth emphasizing that the shape of the variance data is due solely to details of the particle shapes since there is no order in their arrangement.

\subsection{Sub-conclusion}

The above simulation results fully verify the fundamental equation (\ref{hudls1}) for the relative variance ${\mathcal V}(L)=V_R \langle {V_Q}^2\rangle/(V_P {V_\Omega}^2)$ of a random arrangement of extended particles, its special limits, the methods for evaluating it, and the specific predictions it gives for particles of several different common shapes in different dimensions.  The two-dimensional simulations also verify that the effects of total area fraction and polydispersity are understood, and that pixelation effects are important for smaller particles.


\section{Patterns with Hidden Order}

Disordered particle configurations are usually not totally random, but rather possess some degree of order even if it is hidden to the untrained eye.  Here we discuss two ways to quantify this by comparison of measurements with the above predictions for the relative variance of a totally random arrangement of the same particles.  We then analyze simulated patterns of extended particles where some degree of order is induced by excluding the possibility of particle-particle overlaps.  And lastly we analyze Einstein patterns of extended particles where the underlying crystalline order is hidden by displacing each particle by a Gaussian-distributed random length in each dimension.

\subsection{Variance Ratio, $\mathcal R(L)$}

For patterns with hidden order, relative variance measurements ${\mathcal V}_{data}(L)$ must fall below the expectations ${\mathcal V}(L)$ developed above for totally random arrangement of the same set of particles \cite{caveat}.  One way to quantify this is by the volume fraction variance ratio
\begin{equation}
	\mathcal R(L)={\mathcal V}_{data}(L)/\mathcal V(L)
\label{rdef}
\end{equation}
and how it decays with increasing $L$.  If the pattern has long-range density fluctuations, $({\sigma_\phi}^2/\phi)_{data}\sim 1/L^d$, then $\mathcal R(L)$ will decay to a nonzero constant.  By contrast if the pattern is hyperuniform, with $({\sigma_\phi}^2/\phi)_{data}\sim 1/L^{d+\epsilon}$, then it will decay fully to zero with form $\mathcal R(L)\sim 1/L^\epsilon$.  So examining data in terms of $\mathcal R(L)$ removes the effects of dimensionality and allows a yes/no determination of whether or not the system is hyperuniformity, just like just like plots of ${\sigma_\phi}^2(L) L^d$ \cite{DreyfusPRE2015, WuPRE2015}.  But the real-space spectrum $\mathcal R(L)$ additionally removes the effects of particle shape and has meaning in terms of its {\it value}, not just its scaling behavior versus $L$.  In particular, $\mathcal R=1$ means totally random, smaller $\mathcal R$ means more hidden order, and larger $\mathcal R$ means more random.  Thus, $\mathcal R$ can be interpreted as a randomness index.  Another interpretation comes from the special case of a crystalline arrangement of particles where a fraction $f$ of lattice sites are empty.  Such vacancy patterns have Poissonian fluctuations, and the large-$L$ asymptotic value of the variance ratio is calculated to be $\mathcal R(L)=f$ exactly \cite{ATCpixel}.  Thus, $\mathcal R$ can also be interpreted as the fraction of space available for density fluctuations.

\subsection{Hyperuniformity Disorder Length, $h(L)$}

The original idea is that hyperuniform arrangements have fluctuations controlled by the average number of particles on the {\it surface} of the measuring windows.  Then the number variance  scales as surface area, ${\sigma_N}^2(L)\sim \rho L^{d-1}$, and the corresponding volume fraction variance scales as ${\sigma_\phi}^2(L)\sim {\sigma_N}^2(L)/(L^d)^2 \sim 1/L^{d+1}$ \cite{TorquatoPRE2003}.  While $\mathcal R(L)$ is a useful quantity, it does not directly connect to this idea.  So in Ref.~\cite{ATCpixel} we introduced the concept of a hyperuniformity disorder length to make a concrete connection and to give a dimensionally correct form for the scaling of ${\sigma_\phi}^2(L)$.  In particular, for pixel particles and $V_\Omega=L^d$ cubic measuring windows, we defined $h(L)$ such that number fluctuations are given by ${\sigma_N}^2(L)=\overline N_b$ where $\overline N_b=\rho[L^d-(L-2h)^d]$ is the average number of enclosed pixel particles that lie within a distance $h(L)$ of the boundary of the window.  In other words, $h$ distinguishes boundary particles from interior particles, where the latter have average number and standard deviation of $\overline N_i=\rho(L-2h)^d$ and zero.  Then $N_b=N-\overline{N_i}$ is a random variable with mean equal to standard deviation, just as for Poisson statistics (whether or not higher moments of the distribution also satisfy Poisson statistics).  This led to Eq.~(\ref{vpixgen}), which was used for a variety of disordered pixel patterns \cite{ATCpixel}.

To generalize the hyperuniformity disorder length concept for patterns of extended particles, we again define $h(L)$ so that the variance in the number of particles overlapping a set of measuring windows of volume $V_\Omega$ satisfies ${\sigma_N}^2={\sigma_{N_b}}^2=\overline N_b$.  But now $\overline N_b=\rho[V_R]_b$ is given by the volume of a region around the actual $[L^d-(L-2h)^d]$ boundary volume, according to the non-zero size of the particles and how they may partially overlap even though their centers are not enclosed.  The mean-squared overlap between particles and boundary volume is denoted $\langle {V_Q}^2 \rangle_b$, with subscript ``b" for boundary. Then the volume variance is ${\sigma_V}^2={\sigma_N}^2 \langle {V_Q}^2\rangle_b= \rho [V_R \langle {V_Q}^2\rangle ]_b$, and the relative volume fraction variance is $\mathcal V_{data}(L) = ({\sigma_\phi}^2/\phi)_{data}= [ V_R \langle {V_Q}^2 \rangle]_b/(V_P {V_\Omega}^2)$.   Since boundary particles are defined to have ${\sigma_{N_b}}^2=\overline N_b$, as for Poisson statistics, the numerator is $[ V_R \langle {V_Q}^2 \rangle]_b = [V_R \langle {V_Q}^2\rangle]_w-[V_R\langle {V_Q}^2\rangle]_i$, where the two terms are for the {\it w}hole window and for the {\it i}nterior, respectively, also as for Poisson statistics.  For cubic measuring windows, this gives the second fundamental equation of HUDLS as
\begin{equation}
	\boxed{{\mathcal V}_{data}(L)={\mathcal V}(L)-{\mathcal V}(L-2h)\left( \frac{L-2h}{L} \right)^{2d},}
\label{hdef}
\end{equation}
where ${\mathcal V}(L)$ represents the relative volume fraction variance of Eq.~(\ref{hudls1}) for a {\it totally random} arrangement of the same set of particles in the actual pattern.  We emphasize that Eq.~(\ref{hdef}) serves as the definition of $h$.  It can can be rewritten more symmetrically in terms of the variance ratio as
\begin{eqnarray}
	\mathcal R(L) &=& 1 - \frac{ \mathcal V(L-2h)(L-2h)^{2d} }{ \mathcal V(L)(L)^{2d} } \label{hdef2}, \\
	    &=& 1 - \frac{ [\mathcal V {V_\Omega}^2]_i }{ [\mathcal V {V_\Omega}^2]_w},
\end{eqnarray}
where the latter is for measuring windows of arbitrary shape.  Note that for a totally random arrangement, $\mathcal R(L)=1$ and $h(L)=L/2$ hold; these are upper bounds.  Note, also, that for spherical windows Eq.~(\ref{hdef}) becomes ${\mathcal V}_{data}(R)={\mathcal V}(R)-{\mathcal V}(R-h)[(R-h)/R]^{2d}$.

The method of  ``Hyperuniformity Disorder Length Spectroscopy" (HUDLS) is to use the two fundamental Eqs.~(\ref{hudls1},\ref{hdef}) to analyze ${\mathcal V}_{data}(L)$ in terms of the the real-space spectra of $\mathcal R(L)$ and $h(L)$ versus $L$.  Unfortunately, this must be done numerically for most particle shapes because the form of ${\mathcal V}(L)$ computed from Eq.~(\ref{hudls1}) is too complex to be inverted.  But the equations are tractable for large measuring windows with the general limiting behavior ${\mathcal V}(L) \rightarrow \langle H^2\rangle V_P/L^d$ given by Eq.~(\ref{SlargeVB}) for $L^d \gg V_P$.   Then Eq.~(\ref{hdef}) becomes
\begin{eqnarray}
	{\mathcal V}_{data}(L) &=& \frac{\langle H^2\rangle V_P}{L^d}\left\{ 1 - \left[ 1-\frac{2h(L)}{L} \right]^d \right\}, \label{SdatalargeL} \\
	                      &=& 2d\langle H^2\rangle  \frac{V_P h(L)}{L^{d+1}}~~{\rm if}~h(L)\ll L    \label{SdatalargeLsmallH}.
\end{eqnarray}
Recall that $\langle H^2\rangle$ is the mean-squared value of the measuring window hat function; it equals one for step-function windows, in which case the earlier pixel-pattern results are recovered.  Eqs.~(\ref{SdatalargeL}-\ref{SdatalargeLsmallH}) can be inverted for
\begin{eqnarray}
	h(L) &=& \frac{L}{2} - \frac{L}{2} \left[ 1 - \frac{{\mathcal V}_{data}(L)L^d}{\langle H^2\rangle V_P} \right]^{1/d}, \label{hlargeL} \\
	         &=& \frac{ \mathcal V_{data}(L)L^{d+1} }{2d\langle H^2\rangle V_P}~~{\rm if}~h(L)\ll L. \label{hsmall}
\end{eqnarray}
Note that if the measured variance ratio is small, it can thus be interpreted as $\mathcal R(L) = 2dh(L)/L$.

Based on Eqs.~(\ref{hlargeL}-\ref{hsmall}), the following large-$L$ scaling is expected for particles of any shape.  If the pattern has long-range density fluctuations with $\mathcal V_{data}(L)\sim1/L^d$, then $h(L)\propto L$ holds at large $L$.  An example of this is a $d$-dimensional crystal with random vacancies; then $h(L)=fL/(2d)$ where $f$ is the fraction of sites that are vacant \cite{ATCpixel}.  If the pattern is strongly hyperuniform with $\mathcal V_{data}(L)\sim1/L^{d+1}$, then $h(L)=h_e$ becomes constant at large $L$.  An example of this is an Einstein pattern, where particles are independently displaced from crystalline lattice sites as though with thermal energy; then $h_e$ equals about 1/2 the root mean square displacement in each dimension \cite{ATCpixel}.  In general, a large-$L$ scaling of $\mathcal V_{data}(L)\sim1/L^{d+\epsilon}$ implies $h(L)\sim L^{1-\epsilon}$ where $0\le \epsilon \le 1$.

As an aside, pixel patterns with no more than one particle per pixel have $\mathcal V_{data}(p_o)=\mathcal V(p_o)$, as for a random arrangement, because at $L=p_o$ the variance is set by the intensity distribution and does not depend on where the particles happen to be.  The variance ratio of Eq.~(\ref{rdef}) is then given in general by
\begin{eqnarray}
	\mathcal R(L) &\equiv& \frac{\mathcal V_{data}(L) }{\mathcal V(L) } = \frac{\mathcal V_{data}(L)}{\mathcal V_{data}(p_o)}, \\
		&=& 1-\left[1-\frac{2 h(L)}{L} \right]^d,  \\
		&\approx& 2d\frac{h(L)}{L}~~{\rm if}~h\ll L. \label{Rhsmall}
\end{eqnarray}
Variance data for a central pixel representation with good resolution can therefore be normalized to one at $L=p_o$ in order to obtain $\mathcal R(L)$, and this in turn can be simply interpreted in terms of the hyperuniformity disorder length.  It is not necessary to know $\phi$ or $\langle {I_P}^2\rangle/\langle I_P\rangle$ in order to analyze the data.


\section{Demonstrations}

To demonstrate HUDLS we now simulate non-random pixelated patterns and analyze the real-space volume fraction fluctuations in terms of the variance ratio and the hyperuniformity disorder length.   For all examples we use square particles and square $L\times L$ measuring windows, since the variance for random configurations is known exactly from Eqs.~(\ref{S1rectA}-\ref{Sd}).  The particle area is chosen as $(15p_o)^2$, which is just large enough that pixelation effects are not strong and the results are close to continuum.  The system area is at least $(6000p_o)^2$, which is about as large as possible for the current computational resources.  Thus, for a given area fraction, the statistical uncertainty is roughly optimized.

\subsection{Non-Overlapping Particles}

\begin{figure}[ht]
\includegraphics[width=3.000in]{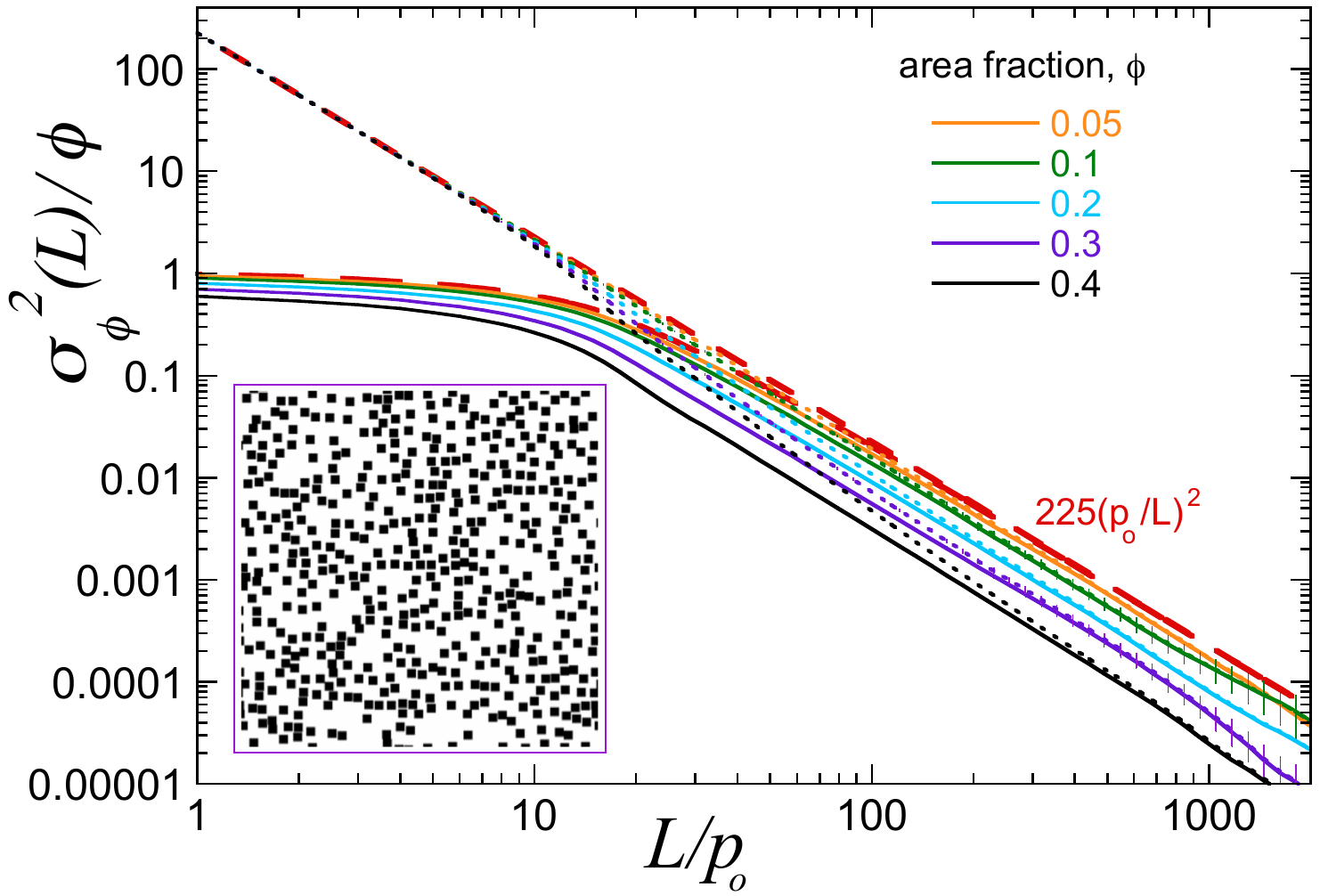}
\caption{Relative variance versus window size for non-overlapping square particles of area $(15p_o)^2$ placed at random into a pixelated image of area $(6000p_o)^2$, with various total area fractions as labeled.  The dotted curves are results for the corresponding central-pixel representation.  The red-dashed curves are the expectations for a totally random arrangement of the same particles.  This is given by $(15p_o/L)^2$ for the central-pixel representation, and by Eqs.~(\ref{S1rectA}-\ref{Sd}) for the actual extended particles.  The inset shows an example pattern for a $(600p_o)^2$ sample with area fraction of 30 percent.  For clarity, error bars are plotted on only two of the data sets.}
\label{nonoV}
\end{figure}

One type of non-random pattern is for particles that are not allowed to overlap, but that are otherwise totally disordered.  For construction, trial locations are chosen at random but are accepted only if the pixels to be covered are all empty.  This is repeated one particle at a time until the desired area fraction is achieved, and gives a binary image of zeros (empty) and ones (covered by one particle).  A small example is shown in the inset of Fig.~\ref{nonoV} for a $(600p_o)^2$ system and 480 square particles of area $(15p_o)^2$, for which the area fraction is 30 percent.  Full-size patterns are created at several area fractions.  Corresponding pixel patterns are simultaneously constructed, where the central pixel for each square particle is set to 225.

The relative variance for five select patterns is plotted versus $L/p_o$ in Fig.~\ref{nonoV}, along with the expectation for random arrangements.  Note that the data appear to approach this upper limit from below as the area fraction is decreased toward zero.  In other words, higher area fractions fall further below the upper bound and are correspondingly less random and have more hidden order.  For large $L$, at any given $\phi$, the variance results for the extended and central-pixel representations merge together, as required, and appear to decay as ${\sigma_\phi}^2(L)\sim 1/L^2$.  This is Poissonian (non-hyperuniform), indicative of long-range density fluctuations as expected for unjammed liquid-like arrangements.  At small $L$, the two representations have very different behavior.  The central-pixel results all collapse onto ${\sigma_\phi}^2(L)/\phi = (15p_o/L)^2$, when viewed on a log-log plot; in fact, the exact behavior is ${\sigma_\phi}^2(L)/\phi = (15p_o/L)^2 - \phi$ for small enough $L$ that all windows have no more than one pixel with non-zero value \cite{ATCpixel}.  By contrast, at small $L$, the variance for the extended-particle patterns become constant: ${\sigma_\phi}^2(L)/\phi \rightarrow {\sigma_\phi}^2(p_o)/\phi = 1-\phi$.  This follows from the expectation ${\sigma_\phi}^2(L)/[\phi(1-\phi)]=(p_o/L)^d$ for random binomial patterns \cite{ATCpixel}, since for $L=p_o$ the variance depends only on the number of zeros and ones in the pattern and not on their arrangement.  The crossover between small- and large-$L$ behaviors happens at about the particle width, as seen in the figure.

\begin{figure}[ht]
\includegraphics[width=3.000in]{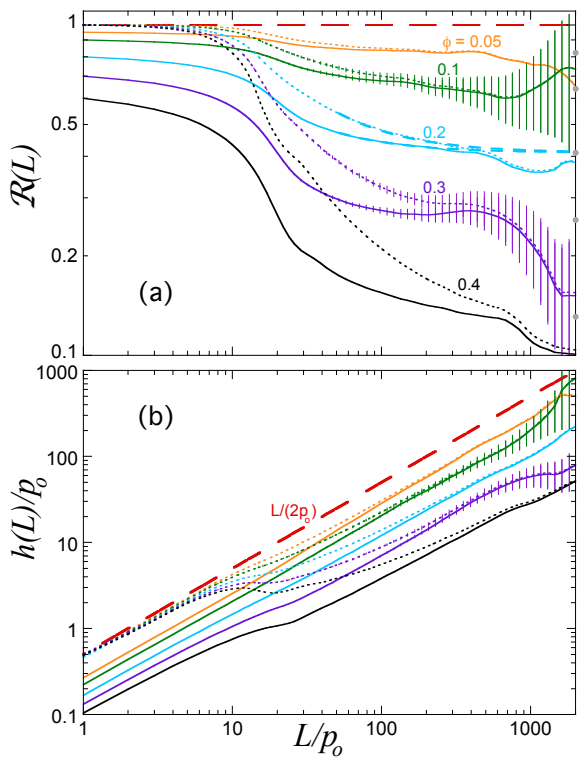}
\caption{(a) Variance ratio and (b) hyperuniformity disorder length based on the results in Fig.~\protect{\ref{nonoV}} for non-overlapping square particles of area $(15p_o)^2$ and labelled area fractions.  For clarity, error bars are plotted on only two of the data sets.  As in Fig.~\protect{\ref{nonoV}}, the solid curves are for the actual extended-particle patterns, and the dotted curves are for the corresponding central-pixel representations.  The curves in (a) are literally the ratio of the Fig.~\protect{\ref{nonoV}} data to the corresponding random-arrangement expectation shown there by the red dashed curves.  The curves in (b) are from solving Eq.~(\ref{hdef}) for $h$.  In (a), the $\phi=0.20$ data at $L>45p_o$ are fit to $R(\infty)[1+ap/L]$, as shown by the blue dashed curves.  Extrapolation results for $\mathcal R(\infty)$ are plotted by solid gray circles on the right y-axis for all five data sets.  Note that the error bars reflect the expected scatter for different patterns of the same size \cite{ATCpixel}, and that this grows at large $L$.
}
\label{nonoRH}
\end{figure}

The same trends can be inspected more easily and critically in Fig.~\ref{nonoRH}a in terms of the variance ratio, $\mathcal R(L)$, defined by Eq.~(\ref{rdef}) as the variance divided by the expectation for a totally random arrangement of the same set of particles.  As observed, the variance ratio starts at $\mathcal R(p_o)=1$ for the central-pixel representations, and at $\mathcal R(p_o)=1-\phi$ for the actual extended-particle patterns.  For $L$ much larger than particle width, it crosses over to the same constant for both representations if the scaling is Poissonian, ${\sigma_\phi}^2(L)\sim 1/L^2$.  The corresponding hyperuniformity disorder lengths are shown underneath, in Fig.~\ref{nonoRH}b, as deduced from Eq.~(\ref{hdef}).  They grow as $h\sim L$ at both small- and large-$L$, but display a developing plateau at about the particle width.  This is more pronounced for the central-pixel representation, since all the data initially behave as $L=h/2$.

Since the  large-$L$ scaling is Poissonian, the simplest description is in terms of the asymptotic large-$L$ value of the variance ratio, $\mathcal R(\infty)$.  For the lowest area fractions, the value of $\mathcal R(\infty)$ can be read right off the graph of $\mathcal R(L)$ versus $L$.  But for large $\phi$, the systems are not big enough for the asymptotic behavior to be fully reached; plus the statistical uncertainty blooms.  Thus a better procedure is to extrapolate by fitting the data for $L>45p_o=3p$ to the form $\mathcal R(L)=\mathcal R(\infty)[1+ap/L]$, where $\mathcal R(\infty)$ and $a$ are adjustable parameters, $p$ is the particle width, and weighting is taken from the expected statistical uncertainty.  This is illustrated in Fig.~\ref{nonoRH}a for the $\phi=0.20$ data.  The fits for both extended and central pixel representations are good, and, importantly, give a consistent value of $\mathcal R(\infty)=0.41\pm0.01$.  The same holds for fits for all the other area fractions, too.

\begin{figure}[ht]
\includegraphics[width=3.000in]{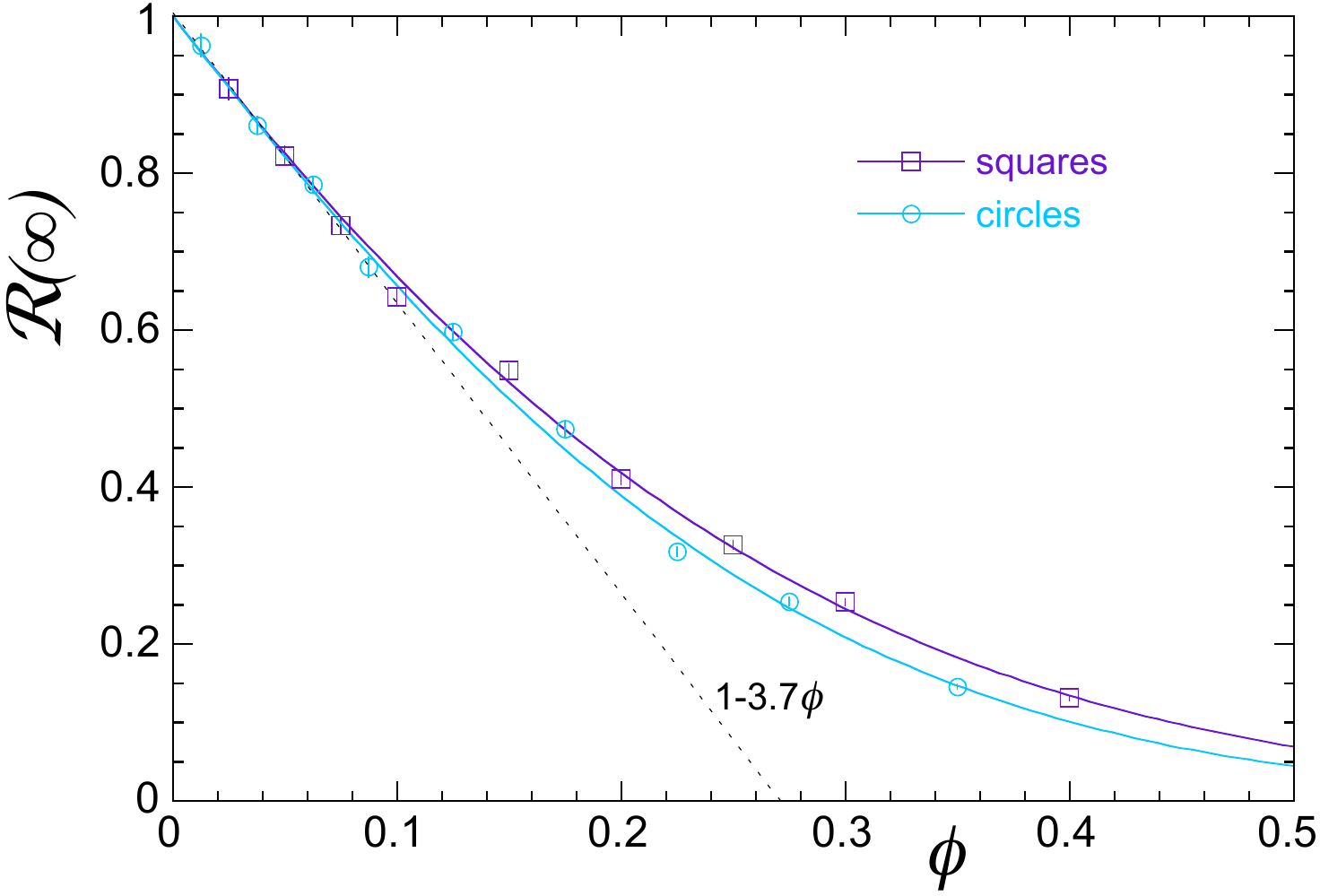}
\caption{Large-$L$ asymptotic limit of the variance ratio versus area fraction.  Values are found by fitting data to the form $\mathcal R(L)=\mathcal R(\infty)[1+ap/L]$ as illustrated in Fig.~\ref{nonoRH}a.  Squares are non-overlapping $(15p_o)^2$ square particles.  Circles are for non-overlapping pixelated circles with the same area.  The solid curves represent a cumulant expansion, $\mathcal R(\infty)=\exp(-3.7\phi-b\phi^2)$, with $b$ equal to 3.3 for squares and 5.1 for circles; the initial behavior is $1-3.7\phi$ as shown by the dashed line.
}
\label{Rinfty}
\end{figure}

Extrapolation results for $\mathcal R(\infty)$ are plotted versus $\phi$ in Fig.~\ref{Rinfty}.  Data are also included for patterns of non-overlapping pixelated circles of the same area as the squares.  As seen, for $\phi<0.1$ the initial behavior closely matches $\mathcal R(\infty)=1-3.7\phi$ for both particle shapes.  For larger $\phi$, the decrease of $\mathcal R(\infty)$ is less rapid, and the circle data fall below the square data.  Presumably this is because $\mathcal R(\infty)$ vanishes, i.e.\ the patterns become hyperuniform, at random-close packing, which depends on particle shape.  This would be interesting to study, but the current construction algorithm is prohibitively slow at larger $\phi$.

\subsection{Einstein Patterns}

\begin{figure}[ht]
\includegraphics[width=3.000in]{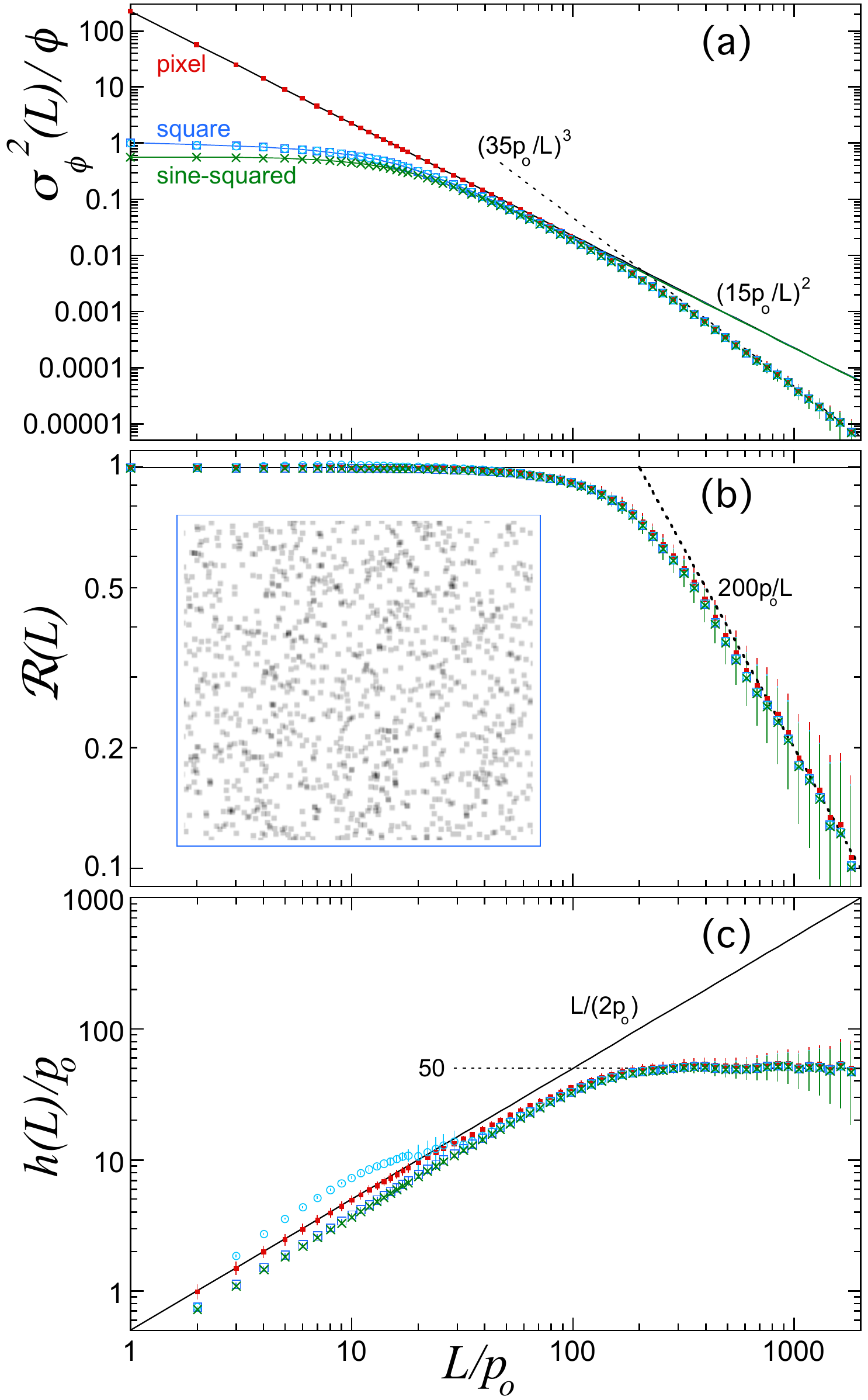}
\caption{(a) Relative variance, (b) variance ratio, and (c) hyperuniformity disorder length versus measuring window size for four Einstein patterns with triangular lattice spacing $b=30p_o$ and root mean square displacement $3b$.  The list of particle centers, the area $a=(15p_o)^2$ of each particle, and the global area fraction $\phi=0.29$, are the same for all four patterns.  The only difference is particle shape: open squares for square particles, solid squares for pixel particles, circles for circular particles, and $\times$s for sine-squared particles.  The solid curves in (a), that match the data at small $L$, are predictions from the text for a random arrangement of particles of the labeled shape; these three curves asymptote to $(15p_o/L)^2$ at large $L$, while the relative variance data all asymptote to $(35p_o/L)^3$.  A small example pattern, for square particles, is shown in (b).
}
\label{Einstein}
\end{figure}

Now we consider a strongly hyperuniform ``Einstein" pattern, where particles are independently displaced from a triangular lattice in $d=2$ dimensions.  Here the lattice spacing is $b=30p_o$, and the displacement distribution is Gaussian with root mean square displacement of $3b$ in each dimension.  In addition to using $(15p_o)^2$ square particles, we also create a corresponding pixel pattern where the central pixel is set to $I_o=225$ for each particle.  And we also create patterns where pixelated circular and sine-squared particles of the same area are placed into exactly the same configuration.   Thus, for all four Einstein pattens, the list of particle centers is the same and the area fractions are all $\phi=I_o\sqrt{4/3}(p_o/b)^2=0.29$.  The only difference is in particle shape.  A small example pattern for the square particles is shown in the inset of Fig.~\ref{Einstein}b; note that it is not binary, since the extended particles can overlap one another.  Note that the underlying lattice is totally hidden to the human eye.

Results for the relative variance are plotted in Fig.~\ref{Einstein}a for all four patterns.  For small $L$ the behavior depends on particle shape, and closely matches the plotted expectations for a random arrangement of the same particles -- for square, pixel, and sine-squared shapes.  For circular particles, the random expectation is unknown but the data are quite close to the square particle results.  For large $L$, the expectation for a random arrangement is ${\sigma_\phi}^2(L)=(15p_o/L)^2$ for all shapes.  As seen, with increasing $L$ the data fall below this limiting power-law but merge together and collapse to approximately ${\sigma_\phi}^2(L)=(35p_o/L)^3$.  Such ${\sigma_\phi}^2(L) \sim 1/L^{d+1}$ scaling is expected for strongly hyperuniform patterns.

This general phenomenology can be seen perhaps more easily in Fig.~\ref{Einstein}b in terms of the ratio $\mathcal R(L)$ of the measured variance to the variance for a random arrangement.  For the circular particles, we divide the variance data by the square particle function.  As such, the variance ratio data nearly collapse together for all four particle shapes.  All data begin very close to $\mathcal R(L)=1$ at small $L$, and cross over to approximately $\mathcal R(L) = (35p_o/L)^3/(15p_o/L)^2 = 200p_o/L$ for large $L$.  The transition between these regimes is set by the root mean square displacement and the value of $3b=90p_o$.

The corresponding hyperuniformity disorder lengths are found from either Eq.~(\ref{hdef}) or (\ref{hdef2}) and plotted in Fig.~\ref{Einstein}c.  At small $L$ they all scale as $h\sim L$ but with a proportionality constant that depends on particle shape.  It is $h=L/2$ for the central pixel representation, as seen in Ref.~\cite{ATCpixel}.  It is about $h=0.37L$ for the square and sine-squared particles; this corresponds to $\mathcal R(L)$ being slightly less than one at small $L$, as expected since the patterns do not have the same intensity distribution as for a fully random pattern.  For circular particles analyzed with the variance function for randomly placed square particles, the behavior of $h(L)$ is a bit irregular at small $L$; it initially matches the square and sine-squared particle results but then rises above the $L/2$ bound.  For large $L$, the hyperuniformity disorder lengths all merge together and approach a constant value as expected for strongly hyperuniform patterns.  The common asymptotic value is about $h(L)=50p_o$, which follows from Eq.~(\ref{Rhsmall}) and $\mathcal R(L)=200p_o/L$; this is slightly larger than half the root mean square displacement in each dimension, as seen for Einstein patterns of pixel particles \cite{ATCpixel}.  The main point of this demonstration is not the particular value, but rather that real-space spectra $\mathcal R(L)$ and $h(L)$ are nearly independent of particle shape by strong contrast with ${\sigma_\phi}^2(L)$.

\section{Conclusions}

In summary we have shown how to compute the volume fraction variance for a totally random arrangement of extended particles, which, by contrast with a point or pixel particle, can lie partially inside and partially outside a measuring window.  And we have shown how this may be used to help quantify hidden order in a disordered pattern either in terms of a variance ratio $\mathcal R(L)$ or a hyperuniformity disorder length $h(L)$.  The former is perhaps more intuitive for liquid-like Poissonian patterns, and the latter for strongly hyperuniform patterns, since the respective quantities become constant for large windows.  Thus we have successfully generalized the HUDLS method of Ref.~\cite{ATCpixel} from pixel or point particles to extended objects.  One benefit is that raw images may now be analyzed directly for the degree of hyperuniformity, as long as the particle size distribution is known, without need for identifying particle positions.  As seen for the non-overlapping particle and Einstein pattern demonstrations, with our new generalizations, both $\mathcal R(L)$ and $h(L)$ become independent of particle shape for large windows.  There features in the real-space spectra reflect only the particle arrangement.  For small windows, however, the spectra reflect both particle shape and spatial arrangement.  These advances will help guide future work on diagnosing hyperuniformity, for example in experiments on foams \cite{ATCfoam} and in simulations of soft discs above and below jamming \cite{ATCjam}.

\begin{acknowledgments}
We thank Jim Sethna for suggesting that the volume fraction variance may be computed from digital image data by a Fourier method, and we thank Steve Teitel for helpful conversations.  This work was supported equally by NASA grant NNX14AM99G and by NSF grant DMR-1305199.
\end{acknowledgments}

%
\bibliography{../HyperRefs}

\begin{thebibliography}{24}
\bibitem[{cav()}]{caveat}
\bibinfo{note}{As a caveat, the variance can actually increase if smaller
  particles aggregate into larger composite particles since
  Eqs.~(\protect{\ref{S1rectA}}-\protect{\ref{S1rectB}}) increase with particle
  width $p$. Thus the variance for a totally random arrangement is an upper
  bound only for repulsive or non-aggregating particles.}
\end{thebibliography}

\end{document}